\documentclass[12pt]{article}
\usepackage{amsmath, amsfonts, amssymb}
\usepackage{graphicx}
\usepackage{enumerate}
\usepackage{acronym}
\usepackage{natbib}
\usepackage{url} % not crucial - just used below for the URL 
\usepackage{color}
\usepackage{caption}
\usepackage{xr}
\usepackage{multirow}
\usepackage{subcaption}
\usepackage{bm}
\usepackage{theorem}
\usepackage{natbib}
\usepackage{acronym}
\usepackage{enumerate}
\usepackage{color}
\usepackage{graphicx}
\usepackage{fullpage} 

\usepackage{tikz}
\usetikzlibrary{shapes.geometric,arrows,positioning, backgrounds}
\tikzstyle{box} = [rectangle, rounded corners, minimum width=1.5cm, minimum height=0.5cm,text centered, draw=black]

\newtheorem{Lemma}{Lemma}[subsection]

\usepackage{tikz}
\usetikzlibrary{shapes.geometric,arrows,positioning, backgrounds,automata,snakes}
\tikzstyle{box} = [rectangle, rounded corners, minimum width=1.5cm, minimum height=0.5cm,text centered, draw=black]

\usepackage[all]{xy}
\usepackage{psfrag}

\usepackage{titlesec}
\titleformat*{\subsection}{\bfseries\itshape}
\titleformat*{\subsubsection}{\itshape}

\newcommand{\be}{\begin{equation}}
\newcommand{\ee}{\end{equation}}
\newcommand{\ba}{\begin{eqnarray}}
\newcommand{\ea}{\end{eqnarray}}
\newcommand{\bi}{\begin{itemize}}
\newcommand{\ei}{\end{itemize}}
\newcommand{\bn}{\begin{enumerate}}
\newcommand{\en}{\end{enumerate}}
\newcommand{\bfi}{\begin{figure}}
\newcommand{\efi}{\end{figure}}

\newcommand{\figfile}{R_Code_and_Figures/Figures}

\def \expectation  {\textrm{E}} 
\def \variance {\textrm{Var} }
\def \covariance {\textrm{Cov} }

\newcommand{\e}[1]{\ensuremath{{\rm E}[#1]}}
\newcommand{\ed}[2]{\ensuremath{{\rm E}_{#1}[#2]}}
\newcommand{\edhat}[2]{\ensuremath{{\hat{\rm E}}_{#1}[#2]}}
\newcommand{\var}[1]{\ensuremath{{\rm Var}[#1]}}
\newcommand{\vard}[2]{\ensuremath{{\rm Var}_{#1}[#2]}}
\newcommand{\vardhat}[2]{\ensuremath{{\hat{\rm V}\rm{ar}}_{#1}[#2]}}

\newcommand{\cov}[2]{\ensuremath{{\rm Cov}\left[#1,#2\right]}}
\newcommand{\covd}[3]{\ensuremath{{\rm Cov}_{#1}\left[#2,#3\right]}}
\newcommand{\aed}[2]{\ensuremath{{\rm \hat{E}}_{#1}[#2]}}
\newcommand{\avard}[2]{\ensuremath{{\rm \hat{V}ar}_{#1}[#2]}}

\newcommand{\eF}[1]{\ed{\bF}{#1}}
\newcommand{\varF}[1]{\vard{\bF}{#1}}
\newcommand{\covF}[2]{\covd{\bF}{#1}{#2}}
\newcommand{\eFX}[1]{\ed{\bF \cup \bX}{#1}}
\newcommand{\varFX}[1]{\vard{\bF \cup \bX}{#1}}

\def \fbexp { \mathbb{E} }
\def \fbvar { \mathbb{V}ar }

\def \identity { \boldsymbol{I} }
\def \independent {\perp\!\!\!\perp} 

\def \normal { \mathcal{N} }
\def \real { \mathbb{R} }

\def \vect {\textrm {vec}}

\newcommand{\ma}{\mathcal{A}}
\newcommand{\mb}{\mathcal{B}}
\newcommand{\md}{\mathcal{D}}

\newcommand{\mx}{\mathcal{X}}
\newcommand{\mz}{\mathcal{Z}}

\def \bc { \mathbf{c} }

\def \bof { \mathbf{f} }
\def \bg { \mathbf{g} }
\def \bh { \mathbf{h} }

\def \bt { \mathbf{t} }
\def \bu { \mathbf{u} }
\def \bv { \boldsymbol{v} }
\def \bw { \boldsymbol{w} }
\def \bx { \mathbf{x} }

\def \bz { \mathbf{z} }

\def \bB { \boldsymbol{B} }
\def \bC { \boldsymbol{C} }
\def \bF { \mathbf{F} }
\def \bG { \boldsymbol{G} }
\def \bH { \mathbf{H} }
\def \bK { \mathbf{K} }
\def \bL { \mathbf{L} }
\def \bT { \mathbf{T} }
\def \bU { \mathbf{U} }
\def \bW { \boldsymbol{W} }
\def \bX { \mathbf{X} }

\def \bb { \bm{\beta} }
\def \bGa{ \bm{\Gamma} }
\def \bDe { \boldsymbol{\mathit{\Delta}} }
\def \bet { \eta }

\def \Sig { \boldsymbol{\mathit{\Sigma}} }
\def \bth { \bm{\theta} }
\def \bxi { \xi }
\def \bTh { \boldsymbol{\mathit{\Theta}} }
\def \bO { \boldsymbol{\mathit{\Omega}} }
\def \bizero { \boldsymbol{\mathit{0}} }
\def \bzero { \mathbf{0} }
\def \mbX { \mathbb{ X } }
\def \mbZ { \mathbb{ Z } }

\def \com { \bh }  % Composite simulator bold h.
\def \cin { \bz } % Composite simulator input bold z.
\def \comin { \com(\cin) } % Composite simulator with input.
\def \simr { \bof } % Component simulator bold f.
\def \siin { \bx } % Component simulator generic input bold x.
\def \simrin { \simr(\siin) } % Component simulator with input. 
\def \simri { \simr^i } 
\def \simrj { \simr^j }
\def \simrcin { \simr^1(\bz) } 
\def \bez { \bt^E(\bz) }
\def \bvz { \bt^V(\bz) }

\def \fx { \simr(\bx) }

\def \cX { \bc(\bX) }

\def \fX { \simr(\bX) }

\def \gX { \bg(\bX) }

\def \uX { \bu(\bX) }
\def \uXp { \bu(\bX') }
\def \vX { \bv(\bX) }

\def \wX { \bw(\bX) }

\def \disp { d }
\def \dr { \rho }
\def \ddr { h }
\def \zSM { z_{SM} }
\def \zWD { z_{WD} }
\def \zWS { z_{WS} }
\def \dispin { \disp( \cin ) }

\def \cl { \theta }
\def \meanvec { \bm{ \mu } }
\def \varmat { \bm{ \nu } }
\def \dist { \pi }

\newcommand{\seq}[2]{#1,\dots,#2}

\newcommand{\seqsup}[3]{\seq{#1^{#2}}{#1^{#3}}}

\def \bOi { \bO^{-1}  } % Omega inverse.
\def \bDi { \bDe^{-1} }
\def \eb { \ed{\bF}{\bb} } % adjusted expectation for beta.
\def \ebt { \ed{\bF}{\bb^T} } % adjusted expectation for beta.
\def \vb { \vard{\bF}{\bb} } % adjusted variance for beta. 
\def \ewX { \e{\wX} } % expectation of w(X)
\def \vwX { \var{\wX} } % variance of w(X)
\def \WDW { \bW \, \bDe \, \bW^T }
\def \WOW { \bW^T \, \bOi \, \bW }
\def \WO { \bW^T \, \bOi }

\newcommand{\NL}{ \nonumber \\ &&  \,\,\,\,\,\, } 
\newcommand{\NLeq}{ \nonumber \\  &=&}

\newcommand{\sam}[1]{\textcolor{black}{#1}}% SEJ comments/changes	
\newcommand{\posapp}[1]{\textcolor{black}{#1}}% SEJ comments/changes	
\newcommand{\dcw}[1]{\textcolor{black}{#1}}% SEJ comments/changes					

\usepackage[refpage]{nomencl}

\makenomenclature

\begin{document}

\title{Bayes Linear Emulation of Simulator Networks}

\author{Samuel E. Jackson\footnote{samuel.e.jackson@ucl.ac.uk}
 \\
\small Clinical Operations Research Unit, University College London, London, UK \vspace{0.3cm} \\
David C. Woods \\
\small Southampton Statistical Sciences Research Institute, University of Southampton,  \\
\small Southampton, UK \\
}

\maketitle

\normalsize
\abstract{Computationally expensive simulators, implementing mathematical models in computer codes, are commonly approximated using statistical emulators.
We develop and assess novel emulation methods for systems best modelled via a chain, series or network of simulators. Using a Bayes linear framework, we link statistical emulators of the component simulators to explicitly account for the simulator input uncertainty induced by links between models in arbitrarily large networks. We demonstrate the advantages of these methods compared to use of a single emulator of the composite simulator network for a variety of examples, including the motivating epidemiological simulator chain to model the impact of an airborne infectious disease.}

%\vspace{-2cm}

%%%%%%%%%%%%%%%%%%%%%%%%%%%%%%%%%%%%%%%%%%%%%%

\section{Introduction \label{intro}}

\setlength{\abovedisplayskip}{3pt}
\setlength{\belowdisplayskip}{4pt}
\allowdisplaybreaks

\vspace{-0.1cm}

%%%NOM%%%
% \nom{\disp}{dispersion simulator}
% \nom{\dr}{dose-response simulator}
% \nom{\com}{composite simulator}
% \nom{\cin}{input to composite simulator $ \com $}

%\begin{itemize}
%\item Computer Models.
%\item Chains and networks of models.
%\item Challenges in the analysis of computer models/requirement for the use of emulators.
%\item Emulator of chain or chain of emulators?  State that we will present two novel approaches for analysing the chain of models by linking their Bayes linear emulators.
%\item Application area: comment briefly - dispersion and dose-response models
%\item Overview of the wonderful stuff to come
%\end{itemize}

\dcw{Scientific processes are commonly modelled using \dcw{mathematical models} implemented in computer codes, or \textit{simulators}, that encapsulate the key features of the system and facilitate prediction and decision making.} \dcw{Complex systems can often be most appropriately modelled as a network of simpler \textit{component} simulators, that together form a \textit{composite} simulator of the entire system of interest. In this paper, we develop statistical emulation methods to facilitate uncertainty quantification for such simulator networks.}  

\begin{figure} 
 \begin{center}
\begin{tikzpicture}[node distance=0.8cm, scale=0.9, every node/.style={scale=0.9}]

    \node [box, text width =2.2cm]              (a1) {wind speed};
    \node [box, below=of a1, text width = 2.2cm] (a2) {wind direction};
    \node [box, below=of a2, text width = 2.2cm] (a3) {source mass};
    \node [box, right=of a2] (a4) {$\cin$};
    \node [box, right=of a4] (a5) {$\disp(\cin)$};
     \node [box, right=of a5] (a8) {$\dr(\disp(\cin))$};
    \node [box, right=of a8, text width = 2cm] (a9) {casualty proportion};
     \draw [->] (a1) -- (a4);
     \draw [->] (a2) -- (a4);
     \draw [->] (a3) -- (a4);
    \draw [->] (a4) -- (a5) node[midway, above] {$\disp$};
     \draw [->] (a5) -- (a8) node[midway, above] {$\dr$};
     \draw [->] (a8) -- (a9);

\end{tikzpicture}
  \caption{Graphical representation of the \sam{DDR} 
 network of simulators $ \ddr( \cin ) = \dr( \disp( \cin ) ) $%, where $ \disp $ represents the dispersion model, and $ \dr $ represents the DR model
 . 
\label{GMDDR}}
\end{center}
\end{figure}
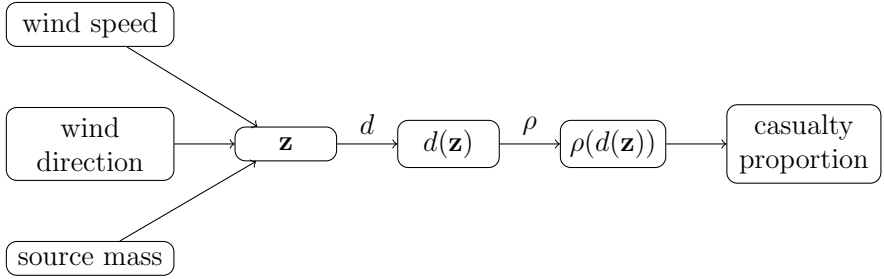

\dcw{One simple, but important, example from epidemiology combines an atmospheric Anthrax dispersion simulator} \sam{\citep{legrand2009etl}}, \dcw{labelled} $ \disp(\cdot) $, \dcw{with} a dose-response (DR) simulator \citep{groer1978drc}, labelled $ \dr(\cdot) $, in a simple \textit{chain} network; \dcw{see Figure~\ref{GMDDR}.} The composite \sam{dispersion dose-response (DDR)} simulator $ \ddr = \dr( \disp( \cin ) ) $ \dcw{models the overall process}, \sam{where $ \cin $ can be viewed as the input to $ \disp $ or $ \ddr $.}

\dcw{In the specific application we consider}, the dispersion simulator models \dcw{the spread of a} released biological agent across a given spatial domain, with input parameters of interest corresponding to physical quantities wind speed ($ \zWS $), wind direction ($ \zWD $) and source mass ($ \zSM $). \dcw{Simulator outputs $ \dispin $ represent dose at each location across the domain}.

For a given spatial location, the DR simulator \dcw{takes dose, $x$, as input} and outputs \dcw{casualties, $ \rho(x)$, as a proportion of the population at that location.} When combined into a modelling chain, we take $x = \disp(\cin)$. That is, the output from the dispersion simulator becomes the input to the DR simulator.

The primary interest of decision makers is the impact of release conditions on numbers of casualties. This assessment requires linking the two component simulators, each of which implements modelling from two different groups of experts. \dcw{Other applications involving chained simulators include modelling of climate \citep{OC5} and seismic activity \citep{CMFP}.}

Utilising \dcw{simulator networks} for \dcw{uncertainty quantification is challenging, largely due to the variety of sources of uncertainty for each individual component simulator \citep{AMA} and the necessity of propagating that uncertainty through the network. In particular, the computational expense of a typical simulator leads to substantial output uncertainty across the input space due to the small number of input combinations for which it is feasible to run the simulator. Often, this uncertainty is captured by building a statistical approximation, or emulator, of the simulator using a computer experiment. In this paper, we answer the important, yet rather under-explored, question of whether combining emulators for the \sam{component} simulators within a network can result in more powerful approximations than emulating the network as a single composite simulator.}
In particular, we present two novel approaches for linking Bayes linear emulators, which extend to arbitrarily large networks of simulators, as well as overcoming certain limitations on the structural form of the emulators seen in related work (for example \cite{CCMLSE} and Section \ref{APSCM}).

\sam{The article is outlined as follows. We formally introduce concepts, notation and previous work for simulator networks in Section~\ref{APSCM}. 
In Section~\ref{AECM}, we present the two novel approaches alluded to above, before
% to obtain an approximation for the composite simulator by linking Bayes linear emulators of several component simulators. 
% These novel approaches are 
demonstrating their efficacy in Section~\ref{ADDR}, using the epidemiological application, by comparison to direct emulation of the composite simulator. 
Motivated by this application, much of the article focuses on the relatively simple composite simulator formed of two \sam{component} simulators, however, in Section~\ref{ENCM} we demonstrate the methodology on a more complicated network of simulators. Section~\ref{conc} contains a brief discussion and some directions for future research. The methods in this paper are implemented in the \texttt{R} package \texttt{NetworkPPBLE} available at \url{https://github.com/Jackson-SE/NetworkPPBLE}.}

%%%%%%%%%%%%%%%%%%%%%%%%%%%%%%%%%%%%%%%%%%%%%%

\section{Networks of Simulators \label{APSCM}}

We represent a simulator network as a directed acyclic graph \citep[DAG;][]{thulasiraman1992gta} with nodes $ \seqsup{\simr}{0}{w} $:
\bn
\item $ \simr^0 = \cin $ is a root note (in-degree of zero), representing a $ p_z $-vector of independent inputs, where $ \cin \in \mbZ \subseteq \real^{p_z} $
% \item $ \al_w = \by $ is an anti-root node (out-degree of zero), representing a $ q_w $-vector of final outputs, 
and
\item $ \simri = \simri(\cdot), i = \seq{1}{w}, $ represent component simulators with generic $p_i$-vector inputs \sam{$ \siin_i \in \mbX_i \subseteq \real^{p_i} $} and $ q_i $-vector outputs.
\en
The edges of the DAG represent directed links between simulators; there is an edge from node $\simri$ to $\simrj$ if an output from simulator $\simri$ is an input to simulator $\simrj$. Further, we order the nodes such that a directed edge can only exist from $\simri$ to $\simrj$ if $i < j$ ($i, j = \seq{1}{w}$).

In general, we can now consider methods of emulating $ \simrj $, given that there is at least one occasion where 
% for at least one input, 
$ x_{j(r)} = f^i_k(\siin_i) $, that is, 
% there is at least one occasion 
where the $r$th input to the $j$th simulator arises as the $k$th output from the $i$th simulator, $i<j$. For any such inputs, note that we also require $ f_k( \mbX_i ) \subseteq \mbX_{j(r)} $, where $ \mbX_{j(r)} $ is the domain for the $r$th input to the $j$th simulator and $ f_k( \mbX_i ) $ is the domain for the $k$th output of the $i$th simulator. For well designed simulator networks, this restriction will be satisfied automatically.

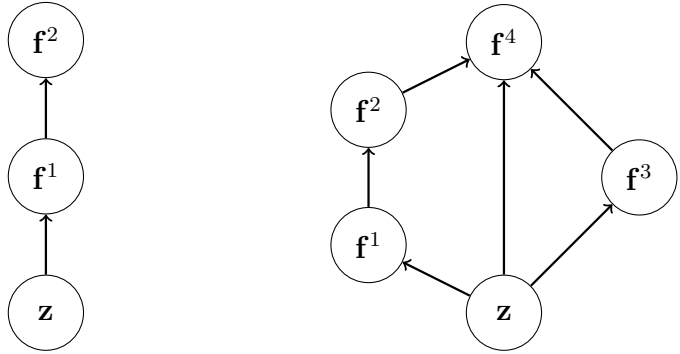
\begin{figure}% [ht]
%\begin{subfigure}{0.4\columnwidth}
\centering
\begin{tikzpicture}[node distance=0.8cm, scale=0.6, every node/.style={scale=1, circle, draw, minimum size = 1cm}]

    \node []              (z) {$\bz$};
    \node [above=of z] (f1) {$\simr^1$};
    \node [above=of f1] (f2) {$\simr^2$};
%    \node [above=of f2] (y) {$\by$};
%    
%
     \draw [->, line  width = 0.3mm] (z) -- (f1);
     \draw [->, line  width = 0.3mm] (f1) -- (f2);
%      \draw [->, line  width = 0.3mm] (f2) -- (y);

\end{tikzpicture}
\hspace{3cm}
%\caption{\footnotesize{Simple chain network.} \label{GRCM}}
%\end{subfigure}
%\begin{subfigure}{0.4\columnwidth}
\begin{tikzpicture}[node distance=0.8cm, scale=0.6, every node/.style={scale=1, circle, draw, minimum size = 1cm}]

    \node []              (z) {$\bz$};
    \node  at (-3,1.5) (f1) {$\simr^1$};
    \node  at (-3,4.5) (f2) {$\simr^2$};
    \node  at (0,6) (f4) {$\simr^4$};
    \node at (3,3) (f3) {$\simr^3$};
%     \node at (0,9) (y) {$\by$};
%    
%
     \draw [->, line  width = 0.3mm] (z) -- (f1);
      \draw [->, line  width = 0.3mm] (z) -- (f4);
     \draw [->, line  width = 0.3mm] (f1) -- (f2);
      \draw [->, line  width = 0.3mm] (f2) -- (f4);
             \draw [->, line  width = 0.3mm] (z) -- (f3);
                    \draw [->, line  width = 0.3mm] (f3) -- (f4);
%           \draw [->, line width = 0.3mm] (f4) -- (y);

\end{tikzpicture}
%\caption{\footnotesize{Complex network.} \label{NEGM}}
%\end{subfigure}
%\resizebox{0.88\textwidth}{!}{\begin{minipage}{\textwidth}
%\vspace{0.3cm}
  \caption{\sam{DAGs representing; \emph{left:} a simple chain network of two simulators, 
 % for example representing the simple network example of Section \ref{subsec:IE}, the example of \cite{CCMLSE} with $ \simr^1 = BENT $ and $ \simr^2 = PUFF $, or the DDR example introduced in Section \ref{intro} and discussed further in Section \ref{ADDR}.  
and \emph{right:} a more complex network of simulators.}
 %, as is discussed in Section \ref{ENCM}. 
 \label{DAGs} }  
%  \end{minipage}}
\end{figure}

Figure~\ref{DAGs} shows two DAGs representing simulator networks.  The left-hand DAG represents a simple chain network of two simulators, which, for clarity, we take as the main focus of the article. Such a chain may represent the simple network example of Section~\ref{subsec:IE}, the example of \citet{CCMLSE} (discussed below), or the DDR example introduced in Section~\ref{intro} and explored in Section~\ref{ADDR}.  The right-hand DAG shows a more complex network of simulators, as is used in Section~\ref{ENCM} to demonstrate the generality of our methods.
%\begin{figure} 
% \begin{center}
%\includegraphics[width=10.2cm]{\figfile/Figure_1}
%  \caption{Graphical representation of a generic chain of simulators $ h(x) = g(\simrin) $.  \label{GRCM}}
%\end{center}
%\end{figure}

When one or more of the individual simulators $\simri$, $i<w$, is computationally expensive and requires the construction of an emulator, the uncertainty arising from the emulation must be propagated through the network. We focus on linking \textit{Bayes linear emulators} for each component simulator and compare to the direct construction of a single emulator for the composite simulator. Essentially, this results in the requirement to build emulators where one or more simulator inputs are uncertain in the computer experiment.  

In previous related work, \dcw{\citet{CCMLSE} proposed coupling two simulators by linking independently developed Gaussian process (GP) emulators of the simulators.  Their motivation arose from potentially having \textit{separate} training runs for two simulators $\simr^1 = $ \emph{bent} and $\simr^2 = $ \emph{puff}, where \emph{bent} simulates volcanic ash plumes arising from a vent and \emph{puff} simulates ash dispersion. As a result, direct emulation (see Section~\ref{DEh}) of the composite simulator defined by the chain is not possible.  For specific GP forms, they derived closed form expressions for the overall mean and variance 
arising from linking the two emulators and applied these quantities within a normal distribution approximation for the composite emulator.  
This availability of second-order posterior statistics for the chain, but the lack of a closed-form distribution, \sam{has inspired us to investigate \sam{(Bayes linear)} approaches to emulation in this context (see Section~\ref{sec:BLE})}. 
% The methods developed in this paper 
In addition, these Bayes linear approaches remove the requirements that the inputs $ \siin_j $ to $\simrj$ must follow a particular distributional form, and that the correlation functions for each component emulator must be power exponential in form, as well as naturally extending to arbitrarily larger networks.}

More broadly, literature exists concerning the problem of 
% GP/BLE 
training 
\sam{emulators of}
% for 
simulators with inputs perturbed by noise, often assuming that the magnitude of the noise is uncertain and typically modelled using one or more additional hyperparameters.
For example, \citet{mchutchon2011gpt} made use of local linear expansions about each input point to allow input noise to be recast as output noise that is proportional to the squared gradient of the GP posterior mean. Under our definition of a simulator network, input uncertainty at a given input is assumed to be represented in the form of a second-order belief specification (arising from the posterior beliefs of a previous emulator).
%known (in the sense that it is formed from our posterior beliefs of a previous emulator).

%As such, the pertinent question is how to efficiently propagate the uncertainty in the inputs through the emulator.
%This is particularly appropriate 
% in the emulator network situation, 
%\sam{for emulating networks of simulators,}
%although additional uses for the methodology are discussed in Section \ref{conc}.

Emulating simulator networks can also be compared to emulation using deep GPs \citep{DGP, HDADGP}.  Deep GPs arise from belief networks about simulator behaviour based on GP mappings, such that layers of GP latent variables exist between simulator input and output, these being marginalised variationally (see, for example, \citealp{titsias2010bgp}).  
% Intermediate nodes act as inputs for the layer below and outputs for the layer above.  Each layer adds model parameters and a regularisation challenge, arising from needing to choose the size of each latent layer.  The latent variables are effectively marginalised out variationally (see, for example, \cite{titsias2010bgp}).
Whilst similar, the intermediate variables in a simulator network represent physical system properties, which aids the construction and modelling
%The primary difference between the latent variables of a deep Gaussian process and the intermediate variables in a simulator network is that, in a network, the variables themselves % between $ x $ and $ h(x) $ (namely $ z = \simrin $) 
% represent physical system properties, these aiding the construction and modelling 
of the emulators for each of the component processes.  
Direct use of a deep GP for the entire network will not exploit this additional information.
%.  This does not mean that the ideas of a deep Gaussian process couldn't be applied 
However, such ideas may be applicable to aid emulation of the \sam{component} simulators, which can then be linked together using the methods that we present here.

%%%%%%%%%%%%%%%%%%%%%%%%%%%%%%%%%

\subsection{Bayes Linear Emulation\label{sec:BLE}}

% We consider a simulator represented by the function $ \simr $,  
In this section we review general Bayes linear emulation methodology, using notation for a generic simulator $ \simr $.
%, which could represent component simulators $ \simri $ or composite simulator $ \com $ in the previous sections. 
The simulator has input vector $ \siin = ( \seq{x_{(1)}}{x_{(p)}} ) \in \mathbb{X} \subseteq \real^p $,  and outputs vector 
$ \simrin = ( \seq{f_1(\bx)}{f_q(\bx)} ) \in \simr(\mathbb{X}) \subseteq \real^q $. 
% We define a Bayes linear emulator to be a meta-model which expresses beliefs 
We represent our beliefs about the behaviour of 
% any scalar simulator output component 
$ \simrin $
% , for any input $ x $, 
in the following form \citep{PFTIMMPS}: 
\be
\simrin = \bg(\siin)^T \bB + \bu(\siin) % = \sum_{v=1}^{m}  t_{v}(\siin) \bb_{.v} + \bu(\siin) 
\, ,  \label{emulator}
\ee
where $ \bg(\siin) $ is an $ m $-vector of known basis regression functions, $ \bB $ is an $ m \times q $ matrix of unknown regression coefficients, and $ \bu(\siin) $ is a $ q $-dimensional weakly-stationary stochastic process.  

Let $\bb = \vect{(\bB)}$ be an $mq$-vector resulting from stacking the columns of $\bB$, with a generic prior specification $ \e{\bb} = \bGa $ and $ \var{\bb} = \bDe $.
% A second-order prior belief specification for $ \simrin $ across $ \mathbb{X} $ is most commonly obtained by assuming 
We also make the common assumptions that $ \e{\bu(\siin)} = \bzero $, $ \cov{\bb}{\bu(\siin)} = \bizero $, 
% with $\bb = \vect{(\bB)}$ a $mq$-vector resulting from stacking the columns of $\bB$, 
and covariance between $ \bu(\siin) $ and $ \bu(\siin') $ is of the form
\be
\covariance[\bu(\siin), \bu(\siin')] = c(\siin,\siin') \, \Sig \label{cov_fct} \,,% = \sigma_{u} c(|x-x'|)  
\ee
for two inputs $ \siin $ and $ \siin' $. Here, $ \Sig $ is a $ q \times q $ output covariance matrix and $ c(\siin,\siin') $ is a stationary correlation function of $ \siin $ and $ \siin' $ \citep{CE, BCCM}; for example, the Gaussian correlation function
\be
c(\siin,\siin') = \exp \left \{ - \sum_{r=1}^p \left ( \frac{x_{(r)} -x'_{(r)}}{\theta_{r}} \right ) ^2 \right \}   \label{GCF} ,
\ee
which depends on the specification of the correlation length parameters $ \theta_{r}, r=1,...,p $.

Suppose $ \bF = \simr(\mx) = ( \seq{\simr_1(\mx)^T}{\simr_q(\mx)^T} )^T $ is an $ nq $-vector with  $ \simr_k(\mx) $ being $n$-vectors of simulator output $ k = \seq{1}{q} $ run at each row of the $ n \times p $ design matrix $ \mx = ( \seq{\bx^{(1)}}{\bx^{(n)}} )^T $.  We can adjust our second-order prior belief specification about $ \simr(\bx) $ across $ \mathcal{X} $ by $ \bF $ using the Bayes linear update equations to obtain posterior quantities:
\be
\eF{\simrin}  =    \e{\simrin} + \cov{\simrin}{\bF} \var{\bF}^{-1} (\bF - \e{\bF})  \label{Exp} 
\ee
\be
\varF{\simrin}  =   \var{\simrin} - \cov{\simrin}{\bF} \var{\bF}^{-1} \cov{\bF}{\simrin}\,. \label{Var}
\ee
See the supplementary material for further details of the Bayes linear approach.

It is worth noting that avoiding unnecessary distributional assumptions is a general advantage of the Bayes linear approach \sam{to statistical inference}.  Inferential statements can still be made, using results such as Chebyshev's inequality \citep{DVM} or Pukelsheim's $ 3 \sigma $ rule \citep{3SR}.

%%%%%%%%%%%%%%%%%%%%%%%%%%%%%%%%%

\subsection{Direct Emulation of Simulator Network (DE) \label{DEh} }

Let $ \com $ represent a simulator network, so that $ \comin $ denotes running the resulting composite simulator starting with initial inputs $ \cin $.
The aim throughout this article is to develop appropriate mean and variance estimators, $ \meanvec_\com(\bz) $ and $ \varmat_\com(\bz) $, for $ \comin $.

% The direct method of emulating $ \com $ 
An obvious estimation approach is Direct Emulation (DE) of $ \com $, which involves applying the Bayes linear update Equations~\eqref{Exp} and~\eqref{Var} directly to $ \com $:
\begin{alignat}{5} 
\meanvec_\com(\bz) & \, = & \,   \ed{\bH}{\comin} \,\, & \, = && \, \, \e{\comin} + \cov{\comin}{\bH} \var{\bH}^{-1} (\bH-\e{\bH}) \label{BLE1h} \\ 
\varmat_\com(\bz) & \, =  & \,  \vard{\bH}{\comin} & \, = && \, \, \variance[\comin] - \cov{\comin}{\bH} \var{\bH}^{-1} \cov{\bH} {\comin} , \label{BLE2h} 
\end{alignat}
%In this case, we have a set of \sam{$ n_z $} training runs
with $ \bH = \com(\mz) $ being $ n_z $ training runs at input locations given by the \sam{$ n_z \times p_z $} design matrix $ \mz $. 
Such emulation neglects the fact that $ \com $ is composed of multiple \sam{component}s, and thus can't take advantage of accuracy gains that can be attained by utilising this fact.  
In addition, the inputs to subsequent simulators must directly correspond to the outputs of proceeding ones.

%%%%%%%%%%%%%%%%%%%%%%%%%%%%%%%%%

\subsection{Illustrative Example \label{subsec:IE} }

We demonstrate DE in the following example, which will be used throughout the article to demonstrate the novel methods for emulating networks of simulators. Consider the three functions\sam{: 
\be
f^1(x_1)  =  0.2x_1 + \cos(x_1) \hspace{0.7cm}
f^2(x_2)  = \exp(x_2/2) - \sin(5x_2) \hspace{0.7cm}
h(z)=  f^2(f^1(z)), \nonumber
\ee
defined over domains of interest $ \mbX_1 = \mbZ =  [0,10] $ and $  \mbX_2 =[-0.5, 2.5] $.} 
These three functions are shown in Figures~\ref{chain_example}a-\ref{chain_example}c. 
Note that by construction \sam{we have $ f^1(\mbX_1) \subset \mbX_2 $.  } In addition, note how chaining even simple functions together can lead to much more complex behaviour.  

Here we focus on emulating $ h $ directly, 
following Equations \eqref{BLE1h} and \eqref{BLE2h},
where training runs $ \bH $ are from $n_z = 8$ equally spaced points over the input domain $\mbZ$ (as shown in Figure \ref{chain_example}c). 

Our prior beliefs about the behaviour of $ h $ are represented by Equation~(\ref{emulator}), with covariance structure given by Equations (\ref{cov_fct}) and (\ref{GCF}) and regression functions $ \bg_h(z) = (1,z)^T $, that is, a first-order polynomial function of $ z $.  
We denote regression and covariance parameters for the emulator for $ h $ by $ \bb_h $, $ \sigma^2_h $ and $ \theta_h $ respectively, where $ \sigma^2 $ represents a generic scalar variance term in place of the covariance matrix $ \Sig $ in Equation \eqref{cov_fct}. We specify vague prior beliefs on $ \bb_h $, leading to posterior mean and variance estimators for $ \bb_h $ equivalent to Generalised Least Squares (GLS) estimators \citep{DPSEHMM}.  

We obtain point estimates for $ \sigma_h^2 $ and $ \theta_h $ via maximum likelihood \citep{ENGPE, Jackson1}. We adjust the prior beliefs for any $ z $ in light of simulator runs $ \bH $ using Equations (\ref{BLE1h}) and (\ref{BLE2h}).
\posapp{Resulting visual diagnostics of DE} are provided in Figure~\ref{chain_example}c, where later they can be easily compared to the results of our proposed approaches, showing simulator behaviour alongside emulator expectation $ \pm 3 $ emulator standard deviations \citep{BLS, DGPE}.  
% We assess the validity of the emulator via standardised prediction errors \citep{BLS, DGPE}
% \be 
% \Lambda_f(x) = \frac{f(x) - \mu_f(x)}{\sqrt{\nu_f(x)}} \nonumber,
% \ee
% with $ \mu_f, \nu_f $ representing mean and variance estimators for generic scalar-output simulator $ f $ with generic input $ x $. 
% Large absolute standardised errors indicate a conflict between the simulator and the (hence invalid) emulator and may arise as a result of prior belief misspecification. 
From Figure~\ref{chain_example}c, the direct emulator for $ h $ is valid, as the simulator output mostly lies within the $ \pm 3 $ standard deviation intervals. However, it is also inaccurate (emulator prediction is far from the true simulator output) and imprecise (large emulator uncertainty).  This motivates exploration of alternative approaches to approximating $ h $, for example by making use of the simpler behaviour present in the component simulators $ f^1 $ and $ f^2 $.

We note at this point that various approaches exist in the literature for obtaining improved emulators for erratic functions, such as $ h $.  Examples include Treed GPs \citep{gramacy2012btg}, local GPs \citep{gramacy2015lgp} and the aforementioned deep GPs, amongst others.  The aim of this article is not to compete with such existing methods, but to demonstrate the efficacy of linking together emulators of the \sam{component} simulators in a network compared with emulating the composite simulator directly.  In particular, these alternative methods, with modification for use in the Bayes linear paradigm, may still be used on any component simulator in a network to improve emulation of that component.

%%%%%%%%%%%%%%%%%%%%%%%%%%%%%%%%%%%%%%%%%%%%%%

\section{Approaches to Emulating Networks of Simulators \label{AECM} }

We now develop methods for predictive inference of $ \comin $, improving on the mean and variance estimates $ \meanvec_\com(\bz) $ and $ \varmat_\com(\bz) $ of Section \ref{DEh}. We focus on the two-simulator chain depicted in Figure~\ref{DAGs} (left), although the methods naturally extend to more complex simulator networks (see Section~\ref{NetworkExample}). We assume that training runs 
$ \bK = \simr^1(\mx^1) $ 
and 
$ \bL = \simr^2(\mx^2) $ are available, at input locations given by the design matrices $ \mx^1 $ and $ \mx^2 $ respectively.
As such, unless $ \mx^2 \subseteq \bK $, we can't apply direct emulation as discussed in Section \ref{DEh}, since $\com (z)$ will not be available for all $z \in \bK$.
A major motivation for developing methods that use emulators of component simulators $\simr^1$ and $\simr^2$ is that each emulator should be cheaper to construct (in comparison to the corresponding direct emulator) by requiring less training points as a result of the component simulators' less complex behaviour. In addition, the fact that the number of training points for $ \simr^1 $ and $ \simr^2 $ need not be the same is beneficial if one simulator is faster to evaluate than the other.  Other benefits are also achieved, as we shall proceed to demonstrate.  

Consider a
% $ \ed{\bF,\bL}{\comin} $ and $ \vard{\bF,\bA}{\comin} $ for any input $ \cin $, 
% calculated via 
sequential adjustment of second-order prior beliefs about $ \comin $ by $ \bK $ then $ \bL $ \citep{BLS}:
\ba
\ed{\bK,\bL}{\comin} & = & \ed{\bK}{\comin} + \covd{\bK}{\comin}{\bL} \vard{\bK}{\bL}^{-1}(\bL-\ed{\bK}{\bL}) \label{BLE1FG} \\ 
\vard{\bK,\bL}{\comin}  & = &  \vard{\bK}{\comin} - \covd{\bK}{\comin}{\bL} \vard{\bK}{\bL}^{-1} \covd{\bK}{\bL}{\comin} .\label{BLE2FG} 
\ea 
Taking $ \meanvec_\com(\bz) = \ed{\bK,\bL}{\comin} $ and $  \varmat_\com(\bz) = \vard{\bK,\bL}{\comin} $ as given by Equation \eqref{BLE1FG} and \eqref{BLE2FG} is impractical since meaningful consideration of the required belief specifications to calculate the expressions on the right-hand side directly is likely to be challenging.
% Equivalently, we could adjust sequentially by $ \bL $ then $ \bK $ by swapping the occurrences of $ \bK $ and $ \bL $ on the right-hand side of Equations (\ref{BLE1FG}) and (\ref{BLE2FG}).  
% Consideration of the required belief specification to calculate \sam{Expressions \eqref{BLE1FG} and \eqref{BLE2FG} directly} is likely to be challenging.  
We therefore make the assumption  that $ \simrcin $ is Bayes linear sufficient for the adjustment of $ \comin $ by  $ \bK $, sometimes written 
\be
\lfloor{ \bK \independent \comin } \rfloor / \simrcin\,  \nonumber
\ee
implying that the training runs $ \bK $ have no effect on our beliefs about $ \comin $ once adjusted by $ \simrcin $. This assumption is analogous to a conditional independence property in the full Bayesian paradigm \citep{jensen1998ait}. As a result
\begin{eqnarray}
\ed{\bK,\bL}{\comin} & = & \ed{\bK}{ \ed{\simrcin,\bL}{\comin} },  \label{KL1}  \\ 
\vard{\bK,\bL}{\comin} & = & \ed{\bK}{ \vard{\simrcin, \bL}{\comin} } +    \vard{\bK}{\ed{\simrcin,\bL}{\comin} }     \label{KL2} ,
\end{eqnarray}
where $ \ed{\simrcin,\bL}{\comin} $ and $ \vard{\simrcin, \bL}{\comin} $ are treated as uncertain quantities, for which we wish to adjust our beliefs in light of $ \bK $.

Formally, the subscripts of $ \ed{\simrcin,\bL}{\comin} $ and $ \vard{\simrcin, \bL}{\comin} $ imply a Bayes linear adjustment of $ \comin $ by $ \simrcin $ and $ \bL $. However, since $ \comin = \simr^2(\simr^1(\cin)) $, it is more appropriate to view these expressions as a standard Bayes linear emulator for simulator $ \simr^2 $ given training runs $ \bL $ assuming the input $ \siin_2 = \simr^1(\cin) $ is known. 
In other words, we slightly approximate (hence the $\hat{\,}$ notation below) the right-hand side of Equations \eqref{KL1} and \eqref{KL2} by focussing on Bayes linear adjustment of simulator outputs and seeking to evaluate:
\begin{eqnarray}
\edhat{\bK,\bL}{\comin} & = & \ed{\bK}{ \bez }  \label{Exp_FG}  \\ 
\vardhat{\bK,\bL}{\comin} & = & \ed{\bK}{ \bvz } +    \vard{\bK}{ \bez }     \label{Var_FG}  ,
\end{eqnarray}
where $ \bez = \ed{\bL}{\comin} $ and $ \bvz =  \vard{\bL}{\comin} $ are now the uncertain quantities we wish to adjust our beliefs about in light of $ \bK $ (uncertain because of their dependence on $ \siin_2 = \simr^1(\cin) $).

From this point there are several approaches to 
% making inferences about $ \comin $ 
obtaining mean and variance estimators $ \meanvec_\com(\bz) $ and $ \varmat_\com(\bz) $ for $ \comin $ given $ \bK $ and $ \bL $. We present three such approaches here. The first is a generalisation of Section \ref{DEh}; the latter two are novel.

%%%%%%%%%%%%%%%%%%%%%%%%%%%%%%%%%

 \subsection{Direct Emulation of Second-Order Belief Specification}

For completeness, we start with a generalisation of the DE approach of Section~\ref{DEh} with the composite simulator split into its constituent parts. In a two-simulator system, a Bayes linear emulator is first constructed for $\simr^2(\cdot) $ using training runs $\bL$. Second, the resulting second-order summaries $\bez$ and $\bvz$ are emulated as functions of $\bz$. Training runs $\bT^E$ and $\bT^V$ for these two emulators are straightforward to obtain from training runs $\bK$ for $\simr^1$, since $ \expectation_\bL[\simr^2(\siin_2)] $ and $ \variance_\bL[\simr^2(\siin_2)] $ can be calculated for any $ \siin_2 = \simr^1(\cdot) \in \bK $ using the Bayes linear emulator for $ \simr^2 $. The resulting expressions for $ \ed{\bK}{\bez}, \vard{\bK}{\bez} $ and $ \ed{\bK}{\bvz} $ can then be used to calculate $ \edhat{\bK,\bL}{\comin} $ and $ \vardhat{\bK,\bL}{\comin} $ via Equations \eqref{Exp_FG} and \eqref{Var_FG}, providing appropriate estimators $ \meanvec_\com(\cin) $ and $ \varmat_\com(\cin) $, respectively.

%Decomposing the composite simulator into $ \simr^1 $ and $ \simr^2 $ before emulating removes the constraint that 
% the second simulator is only evaluated at outputs from the first simulator, that is we are no longer restricted to 
%$ \mx^2 = \bK $,
%that is, evaluation of $ \simr^2(\cdot) $ being restricted to the output evaluations of $ \simr^1(\cdot) $, thus permitting different numbers of training points for each simulator. However,
The overall accuracy of the resulting estimators is limited in different ways by the performance of the separate emulators. The number of points in $\bK$ (equivalently $ \bT^E, \bT^V $) benefits emulation of $\bez $ and $ \bvz $, whereas the performance of $\bez$ as an approximation to $\comin$ depends on the number of points in $\bL$. Hence, for a fixed set of training points $\bK$, an increase in the number of training points for $ \simr^2 $ is unlikely to lead to an improved approximation over the DE method of Section~\ref{DEh} with the same $ \bK $, which effectively has $ \bez = \comin $.

As a result, we compare the developed approaches of this article with DE as introduced in Section \ref{DEh},
% This direct emulator of Section~\ref{DEh} 
which can be recovered by setting $ \mx^2 = \bK $.
%, that is by restricting evaluation of $\simr^2(\cdot)$ to output evaluations of $\simr^1(\cdot)$. 
Then, $ \bT^E = \bH $ and $ \bT^V = \bzero $.  Under the prior assumption $\ed{}{\bvz} = 0$, having $ \bT^V = \bzero $ implies $\ed{\bT^V}{\bvz} = \bzero $ for all $ \cin $. Hence, $ \ed{\bT^E}{\bez} = \ed{\bH}{\comin} $ and $ \vard{\bT^E}{\bez} = \vard{\bH}{\comin} $.  
Combining these results with Equations (\ref{Exp_FG}) and (\ref{Var_FG}) leads to \sam{$ \edhat{\bK,\bL}{\comin} = \ed{\bH}{\comin} $ and $ \vardhat{\bK,\bL}{\comin} = \vard{\bH}{\comin} $}. This is an intuitive special case; it is natural that adjusting our belief specification for $ \comin $ given $ \{ \bK, \bL \} $ with the restriction $ \mx^2 = \bK $ should produce the same results as directly adjusting $\comin$ by $ \bH $.

In the DE approach, the training runs $ \bK $ are only used to calculate 
mean and variance estimates for $ \comin $ via $ \bez $ and $ \bvz $. In Section~\ref{EWUI} we present two novel alternative approaches which more directly use the information we have obtained about $\simr^1$ from $\bK$.

%%%%%%%%%%%%%%%%%%%%%%%%%%%%%%%%%

\subsection{Emulation With Uncertain Inputs \label{EWUI}}

As $ \bK $, the output from the first simulator for design $\mx^1$, only affects \sam{$ \bez $ and $ \bvz $} through $ \simrcin $, under the Bayes linear paradigm we can replace $ \bK $ on the right-hand side of Equations~\eqref{Exp_FG} and~\eqref{Var_FG} with $ \ed{\bK}{\simrcin} $ and $ \vard{\bK}{\simrcin} $. Hence 
\begin{eqnarray}
\edhat{\bK,\bL}{\comin} \label{EKhat1} & = & 
% \ed{\bet^F(\cin), \bL}{\comin} \,\, = \,\,  
\ed{\bet^K(\cin)}{ \bez }    \,, %\label{eq_EV} 
\\
\vardhat{\bK,\bL}{\comin}  & = & \label{EKhat2} 
% \vard{\bet^F(\cin), \bL}{\comin} % \nonumber  \\
% & = & \ed{\bet^F(\cin)}{\vard{\simrcin, \bL}{\comin} } +    \vard{\bet^F(\cin)}{\ed{\simrcin,\bL}{\comin} } \nonumber \\
% & = & 
% \,\, = \,\, 
\ed{\bet^K(\cin)}{ \bvz } +    \vard{\bet^K(\cin)}{ \bez }\,, % \label{eq_VV}
\end{eqnarray}
with $ \bet^K(\cin) = \{ \ed{\bK}{\simrcin}, \vard{\bK}{\simrcin} \} $. 
Obtaining each quantity on the right-hand side of Equations \eqref{EKhat1} and \eqref{EKhat2} is tantamount to 
% $ \edhat{\bK,\bL}{\comin} $ and $ \vardhat{\bK,\bL}{\comin} $ can then be viewed as 
requiring \sam{an adjusted second-order belief specification $ \ed{\bL}{\simr^2(\bX_2)} $, $ \vard{\bL}{\simr^2(\bX_2)} $} for simulator $ \simr^2 $ by training runs $ \bL $, where the input $ \bX_2 = \bof_1(\bz) $ is itself a random variable with a second-order belief specification
$ \e{\bX_2} = \ed{\bK}{\simr^1(\cin)} $ and $ \var{\bX_2} = \vard{\bK}{\simr^1(\cin)} $.
%\sam{Note that such a view is once again more in line with a standard emulation procedure for $ \simr^2 $, rather than being a strict Bayes linear adjustment of emulator output given knowledge of its input $ \bX_2 $.}

From here, we see that this is a special case of requiring 
% $ \simr $ \sam{(above $ \simr^2 $)} to obtain belief statements for 
$ \eF{\fX} $ and $ \varF{\fX} $ for $ \fX $, 
%\bL
%\eF{\fX} & = & \ed{\bxi(\bX)}{\ed{\bX, \bK}{\fX}}, \label{eq_EgZ} \\
%\varF{\fX} & = & \vard{\bxi(\bX)}{\ed{\bX, \bK}{\fX}} + \ed{\bxi(\bX)}{\vard{\bX, \bK}{\fX}},  \label{eq_VargZ}
%\ea
where $ \simr $ (above $ \simr^2 $) is a generic simulator, $ \bF $ (above $\bL $) is a vector of training runs for $ \simr $ at a design matrix of known inputs $ \mx $, and $ \bX $ (above $ \bX_2 $) is a vector of random variables with second-order belief statements $ \bxi(\bX) =  \{ \e{\bX}, \var{\bX} \} $. We therefore present two novel methods to Bayes linear emulation with \sam{random variable} inputs. 
The approach in Section~\ref{BLSA} uses appropriate distributions to integrate $ \bX $ out from the emulator specification, whereas the approach in Section~\ref{UIEA} remains in the Bayes linear paradigm.
For each method, we will state explicit expressions for $ \meanvec_\com(\cin) $ and $ \varmat_\com(\cin) $ as approximations to $ \comin $ for the two-simulator chain discussed above, however, they both generalise directly to arbitrarily large networks of simulators, as discussed in Section \ref{ENCM}.

%%%%%%%%%%%%%%%%%%

\subsubsection{Uncertain Input Sampling (UIS) \label{BLSA} }

For UIS, we assume that random variable $ \bX $ follows an appropriate probability distribution $ \dist(\bx) $ which is consistent with our second-order adjusted belief specification \sam{$ \bxi(\bX) $} for $ \bX $, for example 
\be
\bX \sim \dist(\bx) = \normal(\expectation[\bX], \variance[\bX])  % = \normal(\ed{F}{\simrcin}, \vard{F}{\simrcin})
\nonumber  .
\ee
We then approximate 
$ \eF{\fX} $ and $ \varF{\fX} $ by a fully Bayesian treatment of the expectation and variance (denoted $ \fbexp $ and $ \fbvar $) of $ \eF{\fx} $ and $ \varF{\fx} $ over possible $ \bx $:  
\begin{eqnarray}
\eF{\fX} & \approx & 
% \ed{\bxi(\bX)}{\ed{\bX, \bF}{\fX}} \,\,\, \approx \,\,\, \ed{\dist(\bX)}{\ed{\bX, \bF}{\fX}} \,\,\, = \,\,\,  
\fbexp[\eF{\fx}]  = \int_\mbX \eF{\fx} \, \dist(\bx) \, \mathrm{d}\bx  \nonumber \\
\varF{\fX} 
%& = & \vard{\bxi(\bX)}{\ed{\bX, \bF}{\fX}} + \ed{\bxi(\bX)}{\vard{\bX, \bF}{\fX}} \nonumber \\
%& \approx & 
%\vard{\dist(\bX)}{\ed{\bX, \bF}{\fX}} + \ed{\dist(\bX)}{\vard{\bX, \bF}{\fX}}  \nonumber \\
& \approx &
\fbvar[\eF{\fx}] + \fbexp[\varF{\fx}]\,  \nonumber \\
& = & \int_\mbX \big( ( \eF{\fx} - \fbexp[\eF{\fx}] )^2 + \varF{\fx} \, \big) \dist(\bx) \, \mathrm{d}\bx. \nonumber
\end{eqnarray} 

Although integration with respect to the specified distribution for $\bX$ may be possible for specific distributional choices, in general we propose taking a Monte Carlo approximation, leading to 
% Expressions (\ref{approx_exp}) and (\ref{approx_var}) above, 
% \sam{$ \aed{\bF}{\fX} $ and $ \avard{\bF}{\fX} $}, so that
\ba
\aed{\bF}{\fX} & = & \frac{1}{v} \sum_{k = 1}^{v} \eF{\simr(\siin^{(k)})}, \label{approx_exp_MC} \\
\avard{\bF}{\fX} & = & \frac{1}{v} \sum_{k=1}^{v} ( \eF{\simr(\siin^{(k)})} - \aed{\bF}{\fX} )^2 + \frac{1}{v} \sum_{k = 1}^{v} \varF{\simr(\siin^{(k)})}\,, \label{approx_var_MC}
\ea
%\begin{eqnarray}
%\expectation_{F,G}[h(x)] & \approx & \frac{1}{k} \sum_{i=1}^k \expectation_G[g(z^{(i)})] \label{approx_exp_MC} , \\
%\variance_{F,G}[h(x)] & \approx & \frac{1}{k} \sum_{i=1}^k (\expectation_G[g(z^{(i)})] - \expectation_{F,G}[h(x)])^2 + \frac{1}{k} \sum_{i=1}^k \variance_G[g(z^{(i)})] \label{approx_var_MC} .
%\end{eqnarray}
with $ \siin^{(1)}, \ldots, \siin^{(v)} $ a sample from the distribution of $\bX$. Note that these approximations are based solely on emulator means and variances, and do not require $v$ evaluations of the (potentially expensive) simulator $\simr$.
%Note that such Monte Carlo should be relatively fast as it only involves evaluation of emulators.  

Application of UIS via Equations \eqref{approx_exp_MC} and \eqref{approx_var_MC} given second-order belief specification $ \bxi(\bX) $ results in another second-order belief specification, thus making UIS directly applicable to approximate arbitrarily large networks of simulators.
% To cast Equations \eqref{approx_exp_MC} and \eqref{approx_var_MC} 
In the setting of an emulator network formed from a chain of two simulators, estimators for $ \comin $ are explicitly given by:
\ba
\meanvec_\com(\cin) & = & \frac{1}{v} \sum_{k = 1}^{v} \ed{\bL}{\simr^2(\siin_2^{(k)})}, \label{approx_exp_MC_chain} \\
\varmat_\com(\cin) & = & \frac{1}{v} \sum_{k=1}^{v} ( \ed{\bL}{\simr^2(\siin_2^{(k)})} - \meanvec_\com(\cin) )^2 + \frac{1}{v} \sum_{k = 1}^{v} \vard{\bL}{\simr^2(\siin_2^{(k)})}\,, \label{approx_var_MC_chain}
\ea
where $ \bx_2^{(k)}, k = \seq{1}{v} $ are sampled from a distribution consistent with the adjusted second-order beliefs for $\simr^1(\cin)$, for example, $ \normal( \ed{\bK}{\simr^1(\cin)}, \vard{\bK}{\simr^1(\cin)} ) $.

We can gain insight into the uncertainty arising from emulating the separate \sam{component} simulators by considering the two summations in Equation~\eqref{approx_var_MC_chain}. 
% For example, \sam{consider Equation~\eqref{approx_var}} in the context of the chain depicted in Figure~\ref{GRCM}. 
% each corresponds to one of the two parts of Equation (\ref{approx_var});
% Here, \sam{$ \fbvar[\eF{\fx}] $}
% \approx \variance_F[\expectation_{f(x),G}[g(z]] $ 
The first summation (approximating $ \fbvar[\ed{\bL}{\simr^2(\siin_2)}] $)
reflects uncertainty in $ \comin $ as a result of emulating $ \simr^1 $, and
the second summation (approximating $ \fbexp[\vard{\bL}{\simr^2(\siin_2)}] $)
% \sam{$  \fbexp[\varF{\fx}] $}
% \approx \expectation_F[\variance_{f(x),G}[g(z)]]  $ 
reflects uncertainty in $ \comin $ as a result of emulating $ \simr^2 $.
This separation of contributions to the overall variance of $ \com $ could be insightful for multiple reasons.  
An example is experimental design, 
% briefly discussed in Section \ref{subsubsec:design}, 
where one is allocating computational resource budget between training runs of simulators $ \simr^1 $ and $ \simr^2 $.  Finally, having obtained a Monte Carlo sample, it is trivial to calculate an approximation to \sam{$ \fbvar[\vard{\bL}{\simr^2(\siin_2)}] $}, which may also be useful for design purposes, as well as to construct diagnostic measures.

%%%%%%%%%%%%%%%%%%

\subsubsection*{Illustrative Example \label{Ex1} }

\begin{figure}
\centering
\includegraphics[height=14cm, width = 9.5cm, angle = 270]{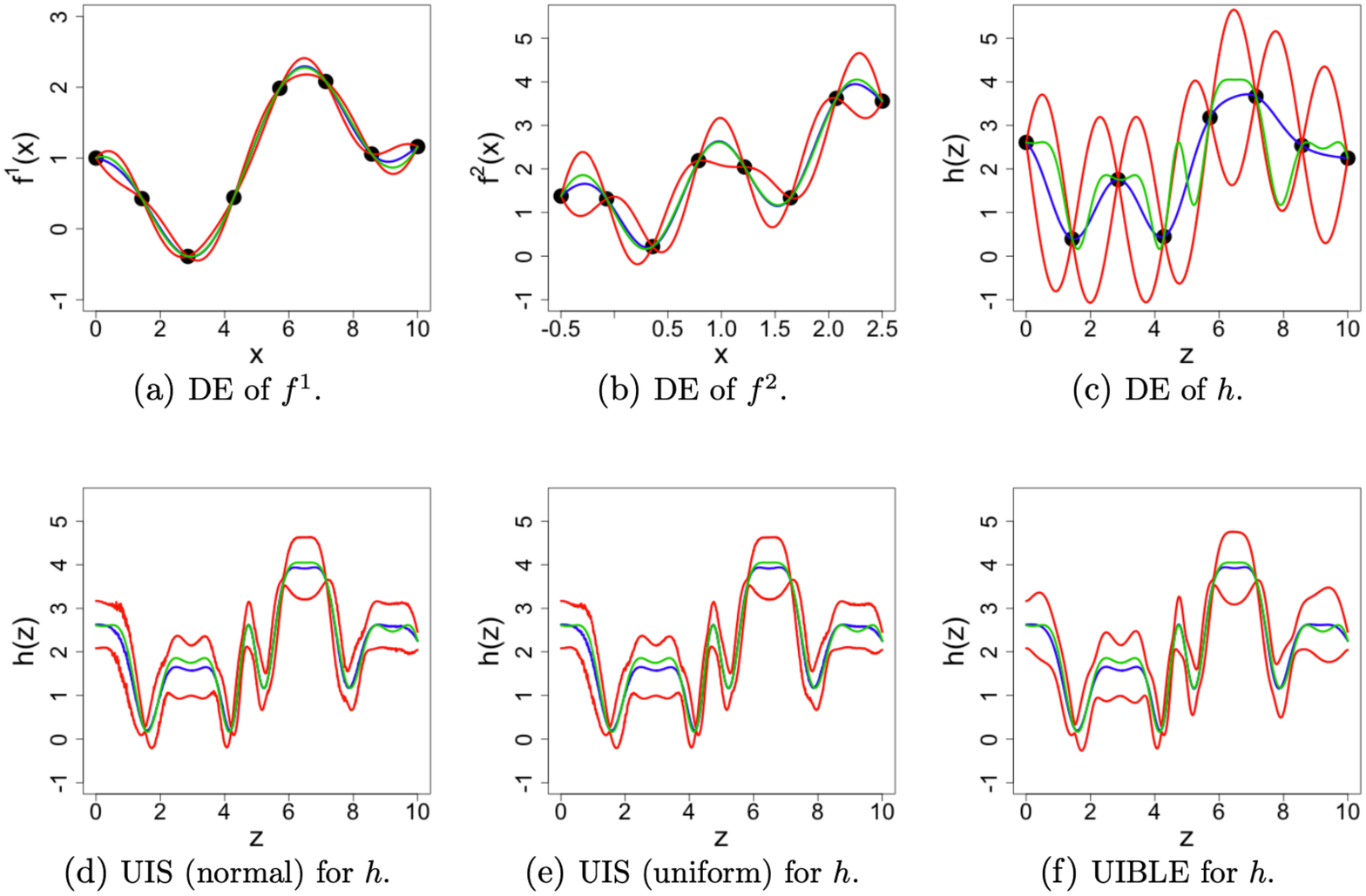}
\caption{Diagnostic plots comparing simulator output (green lines) with 
% various approaches to emulation discussed in Section \ref{AECM}, 
the DE (Section \ref{DEh}), UIS (Section \ref{BLSA}) and UIBLE (Section \ref{UIEA}) approaches to approximating $ h $ shown \sam{as} expectation (blue lines) $ \pm 3 $ standard deviations (red lines). For DE, the training points are represented as black dots. \label{chain_example}}
\end{figure}

%\newcommand{\ChainEgSub}[3]{\subfig{\figfile/Chain_Example/#1}{0.31}{0.3}{-1.5}{#2}{#3}}
%
%\begin{figure}
%\ChainEgSub{DE_of_f_1D}{DE of $ f^1 $.}{DE_1D_f}
%\ChainEgSub{DE_of_g_1D}{DE of $ f^2 $.}{DE_1D_g}
%\ChainEgSub{DE_of_h_1D}{DE of $ h $.}{DE_1D_h}
%\ChainEgSub{ENS_for_h_1D_normal}{UIS (normal) for $ h $.}{ENS_n_1D}
%\ChainEgSub{ENS_for_h_1D_uniform}{UIS (uniform) for $ h $.}{ENS_u_1D}
%\ChainEgSub{UIBLE_for_h_1D}{UIBLE for $ h $.}{UIBLE_1D}
%\vspace{-0.8cm}
%\caption{Diagnostic plots comparing simulator output (green lines) with 
%% various approaches to emulation discussed in Section \ref{AECM}, 
%the DE (Section \ref{DEh}), UIS (Section \ref{BLSA}) and UIBLE (Section \ref{UIEA}) approaches to approximating $ h $ shown \sam{as} expectation (blue lines) $ \pm 3 $ standard deviations (red lines). For DE, the training points are represented as black dots.}
%\label{SampEx}
%\vspace{-0.3cm}
%\end{figure}

Continuing the example of Section~\ref{subsec:IE}, we construct Bayes linear emulators for simulators $ f^1 $ and $ f^2 $ using training sets $ \bK $ and $ \bL $ respectively, each comprising eight simulator runs. We assume the same prior beliefs and techniques for estimating hyperparameters as discussed for $ h $ in \sam{Section~\ref{subsec:IE}}.  The results of emulating these two simulators are shown in Figures~\ref{chain_example}d and~\ref{chain_example}e.  We can see that both of these emulators are valid and accurate, with low uncertainty.
The emulator for $ f^1 $ is more precise, resulting from its simpler behaviour compared to $ f^2 $.

To combine the emulators for $ f^1 $ and $ f^2 $ using UIS we begin by evaluating $ \expectation_\bK[f^1(x_1)] $ and $ \variance_\bK[f^1(x_1)] $ at 1000 evenly-spaced points across the input space for $x_1$. For each value of $ x_1 $, we sampled $ v = 100 $ possible values for $ x_2 = f^1(x_1) $ according to \sam{$ \normal( \ed{\bK}{f^1(x_1)}, \vard{\bK}{f^1(x_1)} ) $},
% a normal distribution with corresponding mean and variance.  We then 
before
% approximated $ \expectation_{\bF,\bA}[h(x)] $ and $ \variance_{\bF,\bA}[h(x)] $ 
\sam{calculating $ \meanvec_\comin $ and  $ \varmat_\comin $}
using Equations (\ref{approx_exp_MC_chain}) and (\ref{approx_var_MC_chain}).  From the diagnostic results of this approximation (Figure \ref{chain_example}d), we observe that $ h $ has been emulated well, with low uncertainty.  Areas of slightly larger uncertainty can be associated with regions of the input spaces for $ f^1 $ and/or $ f^2 $ with larger uncertainty, as should be expected.

To assess the effect of the chosen sampling distribution, we repeat the sampling approximation using a uniform distribution for $ X_2 $ (with parameters chosen such that the first two moments match the second-order belief specification for $ f^1 $ adjusted by $ \bK $).  The results (Figure \ref{chain_example}e) are fairly similar to those assuming a normal distribution, suggesting that the choice of exact distributional specification is not very influential in this case.

%%%%%%%%%%%%%%%%%%

\subsubsection{Uncertain Input Bayes Linear Emulation (UIBLE) \label{UIEA}}

UIBLE is a computationally efficient alternative to UIS, and we profess permits reasonable approximations to simulator networks.  
We consider an emulator setup for $ \simr $ similar to that discussed in Section \ref{sec:BLE}. 
Following Equation \eqref{emulator}, we choose to decompose the vector of training runs $ \bF $ as follows:
\be
\bF = \vect(\bG \bB) + \bU = \bW \bb  + \bU, \nonumber
\ee
where 
$ \bG = ( \seq{ \bg(\siin^{(1)}) }{ \bg(\siin^{(n)})})^T $ is an $ n \times m $ matrix of regressors at the known design points in $ \mx $,  $ \bW = \identity_q \otimes \bG $, $ \identity_q $ is a $ q \times q $ identity matrix, $ \otimes $ represents the kronecker product, $ \bU = \bu(\mx) $ is an $ nq $ vector of residuals, and recall $ \bb = \vect( \bB ) $
with prior specification $ \e{\bb} = \bGa $ and $ \var{\bb} = \bDe $.

We wish to make inference about $ \fX $, where $ \bX \in \real^p $ is an uncertain (random variable) input to $ \simr $.  
Following Equation \eqref{emulator}, $ \fX $ can be written as 
\be
\fX = \gX^T \bB + \uX = \wX \bb + \uX \nonumber.
\ee
where $ \wX = \identity_q \otimes \gX^T $.
We 
assume $ \e{\uX} = \bzero $ and $ \cov{\bb}{\uX} = \mathbf{0} $.  
Such prior specification is similar to one that may be made in the case of known inputs \citep{DPSEHMM}.   
One of the key differences 
is specification of an appropriate correlation function that accounts for \sam{random variable} inputs.  
We assume a general form, similar to that given by Equation (\ref{cov_fct}), as follows:
\be
\cov{\uX}{\uXp} = c(\bX, \bX') \, \Sig = c(\e{\bX}, \var{\bX}, \e{\bX'}, \var{\bX'}, \cov{\bX}{\bX'}) \, \Sig   ,  \label{cov_uXuX}
\ee
which implies that the correlation 
between $ \uX $ and $ \uXp $ for
two uncertain inputs $ \bX $ and $ \bX' $ is 
a function of the second order belief specification about and between the two input variables.
As an example, we propose the following extension to the Gaussian correlation function~(\ref{GCF}):
\be\label{UIcor}
\begin{split}
c(\bX,\bX') & = \exp \left\{ - \e{(\bX - \bX')^T\bTh^{-2}(\bX - \bX')}\right\} \\
 & = \exp \left\{ - \sum_{r=1}^p \left( \frac{\expectation[X_{(r)} - X'_{(r)}]^2 + \variance[X_{(r)} - X'_{(r)}]}{\theta^2_r} \right) \right\} \,,
\end{split}
\ee
with positive-definite diagonal matrix $\bTh^{-2}$ having entries $(\Theta^{-2})_{rr} = 1/\theta_r^2$ and the second line obtained from standard results on the expected value of a quadratic form \citep[][pp. 200-201]{LMRDMA}.  This choice satisfies the desirable property that it reduces to a standard form of correlation function if $ (\bX, \bX') = (\siin, \siin') $ are known.
We also derive two further important results for this new correlation function in the form of two lemmas, proofs of which can be found in the supplementary material.
\begin{Lemma}\label{pdlemma}
For random variables $\bX, \bX'$ with finite first and second moments, the kernel function $c(\bX, \bX')$ from~\eqref{UIcor} is positive-definite.
\end{Lemma}

\begin{Lemma}\label{lblemma}
Covariance function~\eqref{cov_uXuX}, with $c(\bX, \bX')$ given by~\eqref{UIcor}, is a lower bound  under the Loewner (partial) ordering on the covariance obtained by assuming the conditional covariance
\be \label{condcov}
\cov{\uX}{\uXp\,|\,\bX = \bx, \bX'=\bx'} = \exp \left\{ - \sum_{r=1}^p \left( \frac{x_{(r)} - x'_{(r)}}{\theta_r} \right)^2  \right\} \, \Sig\,.
\ee
Moreover, the $kk$th element of~\eqref{cov_uXuX} is a lower bound on the expected value, with respect to $\bX, \bX'$, of the $kk$th element of~\eqref{condcov} ($k=1,\ldots q$).
\end{Lemma}
Whilst the proposed correlation function of Equation \eqref{UIcor} can be viewed simply as a modelling assumption, Lemma \ref{lblemma} shows that it can also be derived as an approximation to the conditional covariance between $ \bu(\bX) $ and $ \bu(\bX') $ assuming the standard Gaussian correlation function that one might use for known $ \bx, \bx' $.  In terms of an emulator, this quantity reflects the amount of resolved uncertainty given the training runs, hence an underestimation (lower bound) of this quantity is preferable to an overestimation.  Similar derivations could be made to extend many other correlation function forms commonly presented in the literature (for example, given by \citealp{DPGP}).

Given 
\sam{the general}
correlation function
\sam{form of Equation \eqref{cov_uXuX} and the}
prior specification above, we have that $ \e{\bU} = \bzero $, $ \var{\bU} = \bO = \Sig \otimes \bC  $ and $ \cov{\bb}{\bU} = \bizero $, where we define
\be
\bC = \left( \begin{array}{cccc} c(\siin_1, \siin_1) & c(\siin_1, \siin_2) & \cdots & c(\siin_1, \siin_n) \\ c(\siin_2, \siin_1) & c(\siin_2, \siin_2) & \cdots & c(\siin_2, \siin_n) \\ \vdots & \vdots & \ddots & \vdots \\ c(\siin_n, \siin_1)& c(\siin_n, \siin_2) & \cdots & c(\siin_n, \siin_n) \end{array} \right)  \nonumber .
\ee
\sam{We also define $ \vX = \Sig \otimes \cX $ and $ \cX = ( \seq{ c(\bX, \siin_1) }{ c(\bX, \siin_n) } ) $.
We now proceed to state the adjusted belief formulae for $ \fX $ by $ \bF $ in the form of two lemmas, proofs of which can be found in the supplementary material. }

\begin{Lemma}\label{lem1} 
The expected value of $ \simr(\bX) $, adjusted by $ \bF $, is given by:
\be
\eF{\fX} = \ewX \, \eb   +    \vX \, \bOi \, (\bF - \bW \, \eb ) \,. \label{UIBLE_lem_E}
\ee
\end{Lemma}
\vspace{-1cm}
\sam{ \begin{Lemma} \label{lem2} 
The variance of $ \simr(\bX) $, adjusted by $ \bF $, is given by:
\ba
\varF{ \fX } & = & \e{ \wX \, \vb \, \wX^T }   +   \ebt \, \vwX \, \eb + \, \Sig  
\NL
\,    - \,   \vX \, \bOi \, \vX^T    +    \vX \, \bOi \, \bW \, \vb  \, \bW^T \, \bOi \, \vX^T
\NL
\, - \, \ewX \, \vb \, \bW \, \bOi \, \vX^T    
\NL
\, - \,   ( \ewX \, \vb \, \bW \, \bOi \, \vX^T ) ^T \,. \label{UIBLE_lem_Var}
\ea
\end{Lemma}  }

\sam{Specification of $  \e{\wX} $ and $ \var{\wX} $} is straight forward for first-order linear regression functions.  
It is also possible for further functions of the input components, but these transformed input components will require a sensible second-order specification.
As for the common known input case, vague priors on $ \bb $ result in $ \eb = \identity_q \otimes ( (\bG^T \, \bC^{-1} \, \bG)^{-1} \, \bG^T  \, \bC^{-1} \, \bF ) $ and $ \vb = \identity_q \otimes ( \bG^T \, \bC^{-1} \, \bG )^{-1}   $.  

The results of Lemmas \ref{lem1} and \ref{lem2} can be used to provide a second-order approximation of the output of any simulator at random variable input for which a second-order belief specification is itself provided.
As a result, UIBLE can be used to approximate arbitrarily large networks of simulators, where the random input $ \bX $ to one simulator is taken to have a second-order belief specification arising from a previous emulator.
In this case, the most straightforward approach to obtaining $  \e{\wX} $ and $ \var{\wX} $ in Equation \eqref{UIBLE_lem_Var} is by emulating the transformed inputs $ \wX $ as further output quantities of the previous simulator.

In the setting of an emulator network formed from a chain of two simulators, estimators for $ \comin $ are explicitly given by application of Equations~\eqref{UIBLE_lem_E} and~\eqref{UIBLE_lem_Var}:
\ba
\meanvec_\com(\cin) & = & \ed{\bL}{\simr^2(\bX_2)} \label{UIBLE_exp_h} \\
\varmat_\com(\cin) & = & \vard{\bL}{\simr^2(\bX_2)} \label{UIBLE_var_h} 
\ea
with $ \e{\bX_2} = \ed{\bK}{\simr^1(\bx)} $ and $ \var{\bX_2} = \vard{\bK}{\simr^1(\bx)} $.

%%%%%%%%%%%%%%%%%%

\subsubsection*{Illustrative Example \label{Ex2} }

We emulate $ f^1 $ at 1000 evenly-spaced points \sam{$ x_1 = z $} across the input space to obtain $ \ed{\bK}{f^1(z)} $ and $ \vard{\bK}{f^1(z)} $.  
The UIBLE for $ f^2 $ is trained using the same training runs as in the previous sections, resulting in the same values for the parameters $ \sigma^2_2 $ and $ \theta_2 $, where we denote by $ \sigma^2_i $ and $ \theta_i $ the scale variance and correlation length parameters for simulator \sam{$ f^i $} respectively.  
Given these parameters, we can now approximate the output to $ h $ at each corresponding uncertain input \sam{$ X_2 $} using Equations \eqref{UIBLE_exp_h} and \eqref{UIBLE_var_h} with adjusted second-order belief specification $ \e{X_2} = \ed{\bL}{f^1(z)} $ and $ \var{X_2} = \vard{\bL}{f^1(z)} $ at each $ z $ of interest.    
The results of doing this are shown in Figure \ref{chain_example}f.

The result of approximating $ h $ using UIBLE is slightly different to that obtained using UIS.  
The blue-line prediction is very similar, however, the $ \pm 3 $ standard deviation bounds are slightly wider in places.  
On the whole, however, we notice that the prediction is quite accurate, with much lower uncertainty than the DE approach \sam{used in Figure \ref{chain_example}c}.

%%%%%%%%%%%%%%%%%%%%%%%%%%%%%%%%%%%%%%%%%%%%%%

\section{Application to a Dispersion Dose-Response Chain of Simulators \label{ADDR} }

In this section, we apply the
% three techniques discussed in Section \ref{AECM} 
% discussed emulation 
UIS and UIBLE methodologies
to the Dispersion Dose-Response (DDR) simulator network introduced in Section~\ref{intro}, comparing these approaches with DE.

Recall that the dispersion model $ \disp(\cdot) $
\citep{brook2003vot} takes input $ \cin $ representing wind speed ($ \zWS $), wind direction ($ \zWD $) and source mass ($ \zSM $), and outputs a biological agent dose $ \dispin $ at a spatial location of interest.
Due to the behaviour of $ \disp(\cdot) $,  we chose to emulate a transformation of the output, namely $ f^1(\siin_1) = \log(\disp(\siin_1) + 1) $, treating this transformed function $ f^1(\cdot) $ as the first simulator of the network.

The DR simulator $ \rho(\cdot) $ takes dose as input and outputs a number of casualties, however, to be consistent with $ f^1(\cdot) $, we consider the second simulator 
to be \sam{$ f^2(x_2)  = \dr( \exp( x_2 ) - 1 ) $}, so that $ h(\cin) = f^2(f^1(\cin)) = \dr( \disp( \cin ) ) $.
We also note that whilst $ \disp(\cdot) $ is computationally expensive, dose-response model $ \rho(\cdot) $ is not; however we emulate both simulators to demonstrate the efficacy of our methods. Our methods are also applicable and effective when only a subset of the simulators in a network require emulation. As $ f^2(\cdot) $ here is straightforward to emulate, our application also effectively demonstrates  the use of the methods for this special case.  
% $ f^2 $ to take input $ \hat{z} = \log(z+1) $ (so that $ f^2(\hat{z}) = \rho(z) $), where $ z $ represents dose.  $ f^2(\hat{z}) $, like $ \rho(z) $, represents number of casualties.  The combined simulator 

The composite simulator $ h = f^2 \cdot f^1 = \rho \cdot d $ takes wind speed, wind direction and source mass as input $ \cin $, and directly outputs a number of casualties $ h(\bz) $.  
The DAG of this setup can be presented as that on the right of Figure \ref{DAGs}, with $ \simr^1 $ and $ \simr^2 $ (now scalar output) as discussed above.
A\sam{n} expanded DAG showing the links between the original simulators $ \disp $ and $ \dr $, their inputs, output and corresponding physical quantities,
%the quantities of interest and the inputs and outputs of simulators $ f $ (dispersion) and $ f^2 $ (DR) 
is presented in Figure \ref{GMDDR}.  

We proceeded to construct Bayes linear emulators for each of the \sam{component} simulators $ f^1 $ and $ f^2 $, as well as the composite simulator $ h $. Ranges of interest of the inputs to simulator $ f^1 $ (and thus $ h $) are 
% as given in Table \ref{IR_f}, 
$ (z_{WD}, z_{WS}, z_{SM} ) \in [37,63]^\circ \times [1,150]\textrm{ms}^{-1} \times [0.001,1]\textrm{kg} $,
each of which were scaled to $ [-1,1] $ for the purposes of our analysis.  We constructed a training point design for $ f^1 $ and $ h $ using a \sam{maximin Latin hypercube} of size 50 across the three input dimensions.  In contrast, simulator $ f^2 $ is one-dimensional, thus the need for fewer training points, so we take a random sample of 20 points from a uniform distribution.  
% Note that, as a result of needing to run simulator $ f^2 $ many times at each training point in order to find the mean function, this simulator is slow, hence if fewer training points are required to train the emulator for this simulator it is already an advantage.  
For each of the emulators for $ f^1,f^2 $ and $ h $, we assumed a Gaussian correlation function, as given by Equation (\ref{GCF}), along with a first-order polynomial mean function.  We represent the scalar variance parameter and correlation length vectors as $ \sigma^2_1, \sigma^2_2, \sigma^2_h $ and $ \bth_1, \cl_2, \bth_h $ respectively. We fit these parameters using maximum likelihood for each emulator, this permitting a fair comparison between the emulation methods presented.  

\begin{figure}
\centering
\includegraphics[height=13.8cm, width = 9cm, angle = 270]{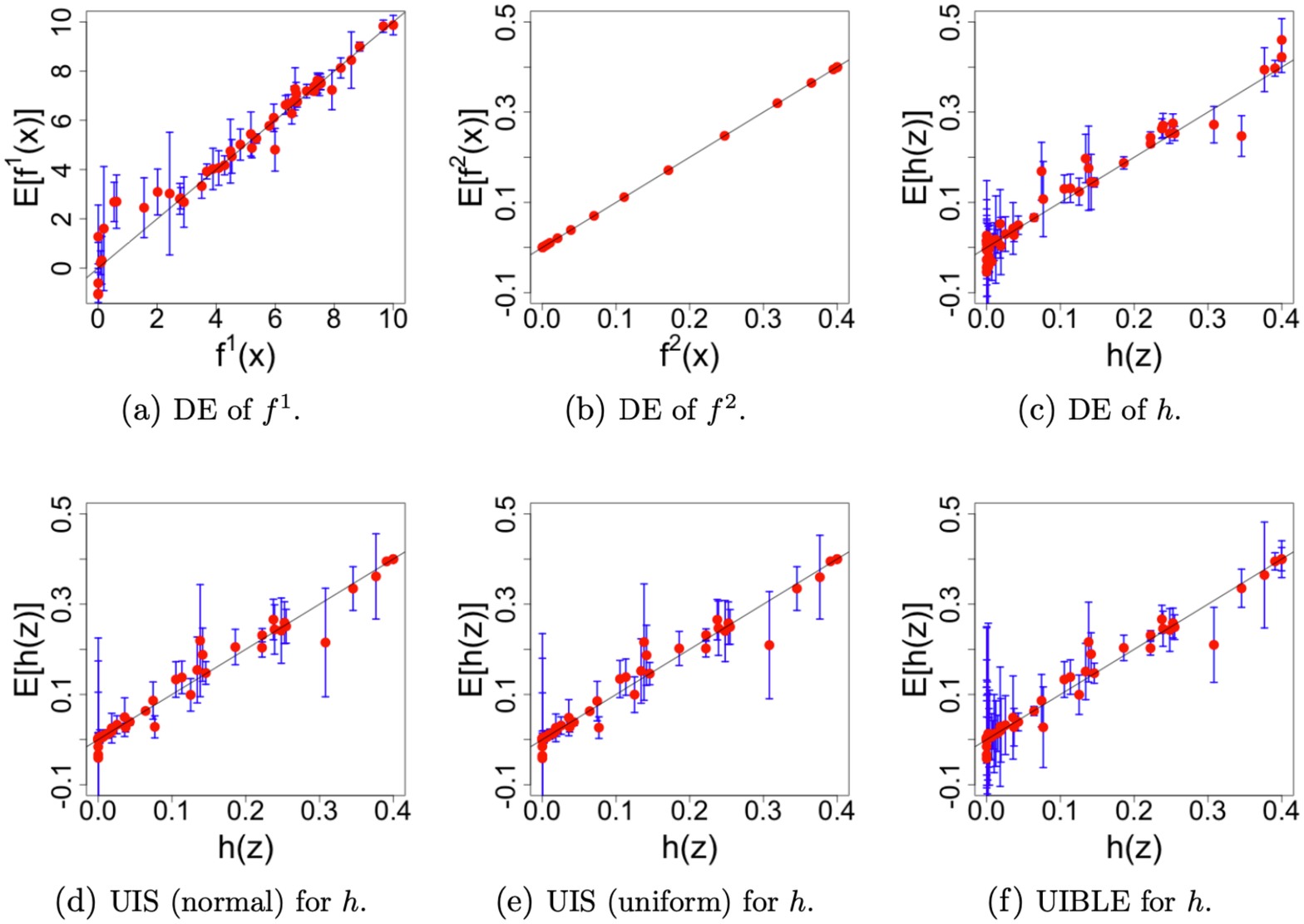}
\caption{Adjusted expectation $ \pm 3 $ standard deviations against simulator output for six different emulators.  $ f^1 $ represents the dispersion simulator, $ f^2 $ the DR simulator, and $ h $ the composite DDR simulator.}
\label{DDR_diagnostics}
\end{figure}

%\newcommand{\DDREgSub}[3]{\subfig{\figfile/DDR_Example/#1}{0.31}{0.3}{-1.2}{#2}{#3}{0.15}}
%
%\begin{figure}
%\begin{center}
%\DDREgSub{DE_for_f_DDR}{DE of $ f^1 $.}{DE_of_Disp}
%\DDREgSub{DE_for_g_DDR}{DE of $ f^2 $.}{DE_of_DR}
%\DDREgSub{DE_for_h_DDR}{DE of $ h $.}{DE_of_DDR}
%\DDREgSub{ENS_for_h_DDR_normal}{UIS (normal) for $ h $.}{ENS_of_DDR_normal}
%\DDREgSub{ENS_for_h_DDR_uniform}{UIS (uniform) for $ h $.}{ENS_of_DDR_uniform}
%\DDREgSub{UIBLE_for_h_DDR}{UIBLE for $ h $.}{UIBLE_of_DDR}
%\end{center}
%\vspace{-0.8cm}
%\caption{Adjusted expectation $ \pm 3 $ standard deviations against simulator output for six different emulators.  $ f^1 $ represents the dispersion simulator, $ f^2 $ the DR simulator, and $ h $ the composite DDR simulator.}
%\label{DDR_diagnostics}
%\vspace{-0.3cm}
%\end{figure}

Given the \sam{component} emulators for $ f^1 $ and $ f^2 $, we can then combine them using UIS and UIBLE to yield chained emulators for $ h $.  Figure \ref{DDR_diagnostics} shows plots of adjusted expectation $ \pm 3 $ standard deviations against simulator output for a set of diagnostic runs for six different approximations; DE of $ f^1 $, $ f^2 $ and $ h $, then approximation of $ h $ via UIS (using normal and uniform sampling distributions) and UIBLE.  
The input \sam{designs} for these diagnostic runs (of size 50 for $ f^1$ and $h $ and 20 for $f^2$) were constructed in the same manner as the training run designs.  
In addition to the plots, Table \ref{RMSE_Tab} shows the Mean Absolute Standardised Prediction Error (MASPE)  \citep{BLS}:
\be
\frac{1}{n} \sum_{k=1}^n \frac{ | f(\siin^{(k)}) - \mu_f(\siin^{(k)}) | }{\sqrt{\nu_f(\siin^{(k)})}}  ,\label{eq_MASPE}
\ee
Root Mean Squared Prediction Error (RMSPE)  \citep{DGPE}:
\be
\sqrt{ \frac{1}{n} \sum_{k=1}^n (f(\siin^{(k)}) - \mu_f(\siin^{(k)}) )^2 }  \label{eq_RMSE} ,
\ee
and Mean Generalised Entropy Score (MGES), as defined by Equation~(27) of \cite{gneiting2007sps}:
\be
- \, \frac{1}{n}\sum_{k=1}^n \left \{ \frac{ \left[ f(\siin^{(k)}) - \mu_f(\siin^{(k)}) \right] ^2 }{\nu_f(\siin^{(k)})} + \log(\nu_f(\siin^{(k)}))  \right \} \label{eq_GR27}
\ee
for the diagnostic runs for each of the six simulators, with 
% $ \psi = f,g $ or $ h $ as appropriate and 
$ \mu_f $, $ \nu_f $ representing appropriate mean and variance estimators corresponding to generic simulator output $ f $.  MASPE is a measure of emulator validity; heuristically we expect this value to be roughly 1 (assuming normal errors this value should be $ \sqrt{2/\pi} $).  RMSPE permits comparison of emulator accuracy.  MGES is larger (better) for approximations that are both valid and precise.

\begin{table}
\caption{The MASPE, RMSPE and MGES for each of the six approximations discussed in Section~\ref{ADDR}. MASPE and RMSPE are smaller-the-better quantities; MGES is larger-the-better. \label{RMSE_Tab}}
  \centering
\fbox{%
\begin{tabular}{c|cc|cccc}
& DE of $f^1$ & DE of $f^2 $ & \,\,\, DE \,\,\, & UIS\_Normal & UIS\_Uniform & \, UIBLE \, \\ \hline
MASPE & 1.638 &  0.766 & 1.579 & 1.256 & 1.242 & 0.767    \\
RMSPE & 0.6457 & 0.0003 & 0.0312 & 0.0235 & 0.0243 & 0.0240 \\
MGES & -1.456 & 14.620 & 4.118 & 7.724 & 7.612 & 6.288 \\
\end{tabular}
}
\end{table}

We can see \sam{from Figure \ref{DDR_diagnostics}a that the} emulator for $ f^1 $ is fairly accurate, with the exception of points towards the bottom end of the output range, where there are several cases of severe overestimation (with underestimated uncertainty).  
The emulator for $ f^2 $ \sam{(Figure \ref{DDR_diagnostics}b)} is very accurate, reflecting the fact that emulator predictions can be taken with almost as much certainty as running the simulator itself. 
As a result, this example serves also to demonstrate the applicability of our methods of approximating simulator networks when only some of the simulators are computationally intensive enough to warrant emulating.
% The diagnostics for these emulators should be kept in mind as we discuss the emulators for $ h $.

The direct emulator for $ h $ \sam{(Figure \ref{DDR_diagnostics}c)} yields predictions with underestimated uncertainty.  
By comparison, the estimated uncertainty for the remaining methods is larger, yielding both more appropriate MASPE values and improved MGES values.  
In addition, the accuracy of the predictions for the chained emulators are, on the whole, improved, this being confirmed by the RMSPE values
% in Table \ref{RMSE_Tab}.  The RMSPE values 
for \sam{UIS} and \sam{UIBLE}.  
It is interesting to note, however, that the uncertainty attributed to each diagnostic point is different between the two approximations, with the uncertainty of \sam{UIBLE} being larger for runs resulting in low or high values of \sam{$ h(\cin) $}, and smaller for those points in the middle.  
This is likely to be a consequence of the way the uncertainty in $ f^1 $ is propagated through $ f^2 $ in the two methods.  
\sam{UIS} propagates uncertainty in $ f^1 $ by sampling possible values of $ f^2 $ according to possible values of $ f^1 $.  
This results in a heteroscedastic error structure across the emulator for $ h $ (for example, if $ f^2 $ is expected to change little regardless of the possible values of $ f^1 $, the uncertainty is small).   
In contrast, \sam{UIBLE} has uncertainty from the regression part and covariance structure.  
As with standard Bayes linear emulation that uses a single correlation structure across $ \mathbb{X} $ 
% (or in this case $ \expectatio $ and $ \variance [\mathcal{X}] $)
, there is  some averaging of the uncertainty estimates for simulator prediction across the input space, even if the behaviour at some points is smoother than others.  
Incorporation of more sophisticated methodology into \sam{the UIBLE} methodology, for example, utilising similar ideas to local GPs \citep{gramacy2015lgp}, may be of benefit in this case.  
To summarise, we feel that the results presented give evidence for the two methods presented for linking emulators \sam{of component simulators in a network} over \sam{using a direct} emulator \sam{of the} composite simulator in many cases.  
We defer further discussion to Section \ref{conc}.

%%%%%%%%%%%%%%%%%%%%%%%%%%%%%%%%%%%%%%%%%%%%%%

\section{Application to a Larger Simulator Network}
\label{ENCM}

Both UIS and UIBLE directly generalise to more complex networks of simulators by repeated application of the general results \eqref{approx_exp_MC}, \eqref{approx_var_MC} or \eqref{UIBLE_lem_E}, \eqref{UIBLE_lem_Var} respectively. 
Such application is possible since the second-order specification
% where the second-order specification 
resulting from application of UIS or UIBLE to one simulator leads to the sampling distribution (for UIS) or uncertain inputs specification (for UIBLE) of the next one.

\begin{figure}
\centering
\includegraphics[height=13.8cm, width = 9cm, angle = 270]{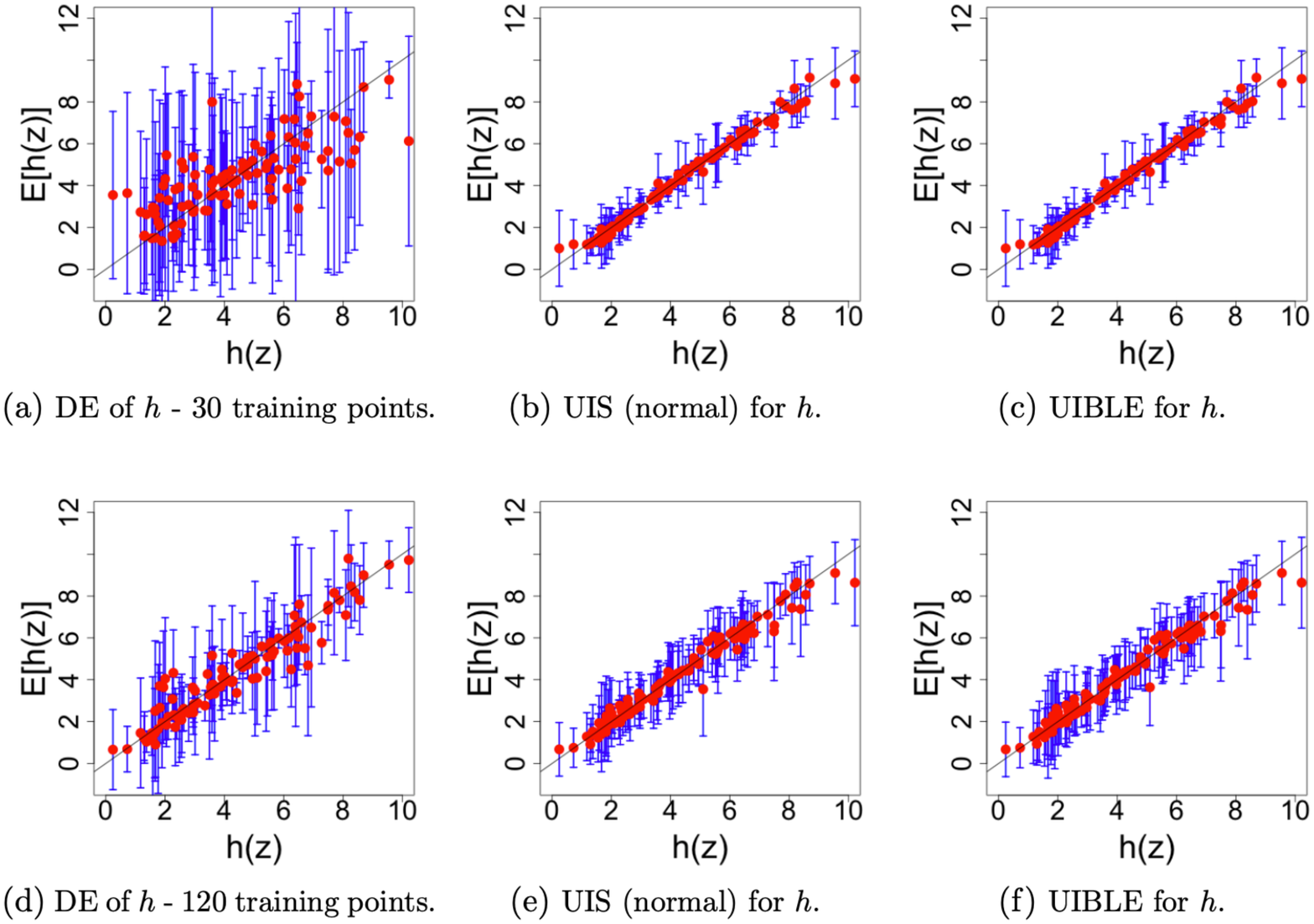}
\caption{Adjusted expectation $ \pm 3 $ standard deviations against simulator output for six different emulators, \sam{as discussed in the text}.}
\label{NetworkExample}
\end{figure}

%\subsection{Illustrative Example \label{ENCMTE} }
% We here demonstrate the applicability of our methods to a more complicated network of simulators.  We analyse the small network of simulators shown in Figure \ref{NEGM}.  
In this section, we consider the illustrative network of simulators shown in Figure \ref{NetworkExample}.  
$ f^1 $ and $ f^2 $ are \sam{taken to be} the same functions as in Section \ref{subsec:IE}, with $ f^3 $ and $ f^4 $ being defined \sam{as follows:
\be
f^3(x_3) = \sqrt{|x_3^3|} - 1.6^{x_3},   \hspace{1.5cm}
%\ee
%with domain of interest $ y \in [-4,6]  $ and $ l $ is the three-dimensional function
%\be
f^4(\siin) = x_{4(1)}x_{4(3)} + \frac{x_{4(2)}}{x_{4(3)}} + \cos(x_{4(1)} + x_{4(2)}) \nonumber  ,
\ee
with $ \mbX_3 = [-4,6]  $ and $ \mbX_4 = [0,4] \times [-2,8] \times [1,2.5] $ (deliberately constructed to 
% satisfy $ f(\mbX_i) \subset \mbX_j $
contain the relevant output domains of previous simulators)}.
The network function $ h $ is defined by:
$$ h(\bz) = f^4\big( f^2( f^1(z_1) ), f^3(z_2), z_3 \big) $$ 
with input \sam{$ \bz = (z_1, z_2, z_3) = (x_1,x_3,x_{4(3)}) $}.  

To begin with, we construct Bayes linear emulators for $ f^i, i = \seq{1}{4} $, and $ h $.   
We take the training points for $ h $ to be a Latin hypercube of size 30 across the three dimensions, appealing to the rough heuristic suggesting a minimum of $ 10p $ design points, where $ p $ is the parameter space dimension \citep{CSSCE}.  
The relevant inputs of this Latin hypercube can then also be used as the training points for $ f^1 $ and $ f^3 $.
% For $ g $, a new set of 30 equally spaced points are taken as the training points $ z $.  A Latin hypercube of size 30 was taken as the training points $ w $ for the emulator for $ l $.  
For $ f^2 $ and $ f^4 $, additional training sets of 30 points were used.
We again assume emulators of the form given by Equation (\ref{emulator}) with covariance structure given by Equation (\ref{GCF}).  
We specify vague prior beliefs on $ \bb $, fitting $ \sigma^2 $ and $ \bth $ by maximum likelihood.  
The emulators for $ f^1 $, $ f^2 $, $ f^3 $ and $ f^4 $ were then combined similarly to the previous examples using both \sam{UIS} (\sam{sampling from} Normal distributions) and \sam{UIBLE} to yield approximations for $ h $, these being compared with DE of $ h $ via the diagnostic plots shown in Figures \ref{NetworkExample}a-\ref{NetworkExample}c.  The design for the diagnostic points was taken to be a Latin hypercube of size 100 across the three input dimensions to $ h $.
%The results of these six emulators are shown in Figure \ref{NE1}.

%\newcommand{\NetEgSub}[3]{\subfig{\figfile/Network_Example/#1}{0.31}{0.3}{-1.2}{#2}{#3}{0.15}}
%
%\begin{figure}
%\begin{center}
%\NetEgSub{DE_of_k_30}{DE of $ h $ - 30 training points.}{Main-DE_of_k_30}
%\NetEgSub{ENS_for_k_30_normal}{UIS (normal) for $ h $.}{ENS_for_k_30}
%\NetEgSub{UIBLE_for_k_30}{UIBLE for $ h $.}{Main-UIBLE_for_k_30}
%\NetEgSub{DE_of_k_120}{DE of $ h $ - 120 training points.}{Main-DE_of_k_120}
%\NetEgSub{ENS_for_k_mixed_normal}{UIS (normal) for $ h $.}{ENS_for_k_mixed}
%\NetEgSub{UIBLE_for_k_mixed}{UIBLE for $ h $.}{Main-UIBLE_for_k_mixed}
%\end{center}
%\vspace{-0.8cm}
%\caption{Adjusted expectation $ \pm 3 $ standard deviations against simulator output for six different emulators, \sam{as discussed in the text}.}
%\label{NetworkExample}
%\vspace{-0.3cm}
%\end{figure}

The behaviour of $ h $ is hard to mimic using DE, whereas  
% Both the sampling approach and uncertain input emulators perform well for $ h_1 = g \cdot f $ with this many points due to the accuracy of the individual emulators.  
UIS and UIBLE yield much more accurate, and very similar, approximations.
% than the direct emulator.  The most obvious benefit of emulating the individual simulators in this example, as for many networks of complex simulators, arises from their reduced dimension.  Since many of the individual simulators are 1-dimensional, their behaviour can be captured accurately.
This is a result of the accuracy to which the component simulators can be emulated, arising largely from their reduced dimension.
Diagnostic plots for the component emulators can be found in the supplementary material, along with further discussion.

Figures \ref{NetworkExample}d-\ref{NetworkExample}f show a repeat of the analysis shown in the top row having increased the number of training points for DE of $ h $ to 120, whilst reducing the number of training points for the emulators of $ f^1 $, $ f^2 $ and $ f^3 $ to 8 (as was the case in Sections \ref{Ex1} and \ref{Ex2}).  
% Due to the reduction in training point number, we this time fitted $ \sigma^2 $ and $ \theta $ using LOOCV, as recommended in \cite{DACE} for emulating simulators using small numbers of training points.  
All other aspects of emulator construction remained the same. 
DE of $ h $ using 120 training points is much more accurate than that using 30, however, only similarly accurate to UIS and UIBLE using many fewer training points.
%  to construct the emulators for which diagnostics are shown in the bottom row of Figure \ref{NE2}.  
Whilst again providing evidence \sam{of} the advantages of \sam{UIS and UIBLE} over DE, there seems to be little discrepancy between the\sam{se} two proposed approaches, with the approximations being very similar.  This is in contrast to the application example of Section \ref{ADDR}, for which UIS and UIBLE yielded different, though comparably valid and accurate, results.
Table \ref{NetTab} shows the MASPE, RMSPE and MGES (as given by Equations \eqref{eq_MASPE}, \eqref{eq_RMSE} and \eqref{eq_GR27} respectively) for each of the six approximations discussed above, with the numbers in brackets indicating the number of training points used to construct the emulators of the simulators in the network.  These confirm the visual diagnostics presented in Figure \ref{NetworkExample}.  Whilst the MASPE values for UIS and UIBLE with 8 and 30 points may be a little low, we note that this slight overestimation of the uncertainty is preferable to underestimation in the context of emulation.
% Further discussion of the results of this example is presented in Appendix \ref{ExtNetEg} of the supplementary material.

\begin{table}
\caption{The MASPE, RMSPE and MGES for each of the six approximations discussed in Section \ref{ENCM}.  % The numbers in the brackets indicate the number of training points used to construct the emulators of the simulators in the network.  
\label{NetTab}}
  \centering
	\begin{tabular}{|c|ccc|ccc|} \hline
		 & DE (30) & UIS (30) & UIBLE (30) & DE (120) & UIS (8, 30) & UIBLE (8, 30)  \\ \hline
			MASPE & 1.105 &  0.748 & 0.749 & 1.056 & 0.662 & 0.627    \\
		RMSPE & 1.512 & 0.260 & 0.260 & 0.839 & 0.406 & 0.407 \\ 
		MGES & -19.274 & 1.967 & 1.965 & -0.446 & 0.452 & 0.385 \\ \hline
	\end{tabular}
\end{table}

%%%%%%%%%%%%%%%%%%%%%%%%%%%%%%%%%%%%%%%%%%%%%%

\section{Discussion and Closing Remarks \label{conc} }

We have presented novel methodology for efficient emulation of networks of simulators. Our examples have shown that both UIS and UIBLE can result in more accurate approximations compared to DE of the composite simulator of the network. 

Each of the demonstrated approaches may be more applicable in different situations.
% Both the sampling approach and uncertain input approach emulators were comparable 
% In addition, some users may feel more comfortable with the 
UIS utilises distributional modelling assumptions for sampling purposes and thus more closely approximates a fully Bayesian analysis. 
Note that the Bayes linear framework in which this paper is largely set does not prevent the investigation of the consequences of assuming certain distributions. 
However, the sensitivity of results to the choice of sampling distribution should be explored by performing a robustness analysis. 
The sampling nature of UIS inherently requires running a standard emulator \sam{many times (at different points)} for a single evaluation. If it is required to evaluate $ \fX $ at very many points, then the approximations~\eqref{approx_exp_MC} and~\eqref{approx_var_MC} could themselves be emulated using a stochastic simulator \citep{PSEM, RESDSSE}. This would avoid the need to approximate $ \simr $ multiple times for each $ \bX $ specification.

In contrast, UIBLE is computationally more efficient as a result of each evaluation being akin to a single run of a standard emulator. The modelling assumptions (particularly regarding the correlation function form) are pragmatic but the resulting emulator can, and should, be assessed using diagnostic summaries and plots. Overall, in the examples presented, no large differences in predictive ability were found between the two methods, although in Section~\ref{ADDR} the two approaches showed different levels of accuracy across different parts of the input space.  
%In addition, the uncertain input emulator is much more efficient due to the fact that the sampling approach requires running the emulators at many more points as a result of sampling, although this will most likely only be noticable for simulators with large numbers of training runs.  In the examples discussed this was not an issue, and run times were fast on a standard laptop computer.

% Whilst we feel that the presented methodology is particularly applicable in situations involving networks of emulators, it should also be noted that UIBLEs have a more general role to play in the emulation of simulators with uncertain inputs.   
In addition to situations involving networks of simulators, the uncertain input emulation approaches discussed here have more general application. For example, they would permit efficient sensitivity analyses; several evaluations of an emulator with constant $ \expectation[\bX] $ and varying $ \variance[\bX] $ \sam{could} quickly provide an indication of the influence of individual inputs on simulator output behaviour.
% In addition, due to the structure of the correlation function, slight modification of the update equations presented in Section \ref{UIEA} would permit the model runs to have arisen from uncertain inputs as well, which may be useful in some situations.

There are multiple directions for future work, for example by developing the methodology to allow for stochastic simulators and ensembles of competing simulators where model selection is required. Interesting design questions arise where several simulators are linked together. In particular, the efficiency of running the various simulators may vary, as might the number of training points deemed appropriate to capture simulator behaviour to a reasonable degree. Deciding how to allocate a fixed computational budget across computer experiments for the individual simulators is therefore an important follow on from this work.
%for example, multiple runs of one simulator may involve a similar amount of computational intensity as a single run of another.  As an example, the dose-response simulator of Section \ref{ADDR} was computationally cheap relative to the dispersion simulator.  Whilst in this case the dose-response model was both faster and simpler, it may be that some component simulators of a network are computationally heavy, whilst exhibiting relatively simple behaviour when analysed in isolation to the entire chain.  
As a final thought, 
% there is scope for future research in the area of parameter estimation for chains of emulators.  
note that in this article we constructed the \sam{component} emulators (including estimation of parameters) 
% such that they satisfied diagnostics 
before combining them together.  However, 
% it may be more beneficial to utilise 
a combined parameter estimation process over all of the \sam{component} emulators of the simulators in a network may prove to be a highly valuable addition to this research.

%%%%%%%%%%%%%%%%%%%%%%%%%%%%%%%%%

\subsection*{Acknowledgements}

% This work was undertaken as part of the ``Crystalcast'' project funded by the US Defence Threat Reduction Agency. 
This work was supported by Chemical and Biological Technologies Department (contract HDTRA1-17-C-0028).
We are grateful to Crystalcast project members for invaluable discussions, comments, and provision of the simulators for the dispersion dose-response application. Particular thanks are due to Professor Veronica Bowman and Dr Daniel Silk (Defence Science and Technology Laboratory, UK), and Dr Daria Semochkina (University of Southampton, UK).

%%%%%%%%%%%%%%%%%%%%%%%%%%%%%%%%%%%%%%%%%%%%%%

\bibliographystyle{agsm}

\bibliography{Biblionotes}

@article{brook2003vot,
	Author = {Brook, D. R. and Beck, N. V. and Clem, C. M. and Strickland, D. C. and Griffiths, I. H. and Hall, D. J. and Kingdon, R. D. and Hargrave, J. M.},
	Journal = {International Conference on Harmonisation within Atmospheric Dispersion Modelling for Regulatory Purposes},
	Pages = {8-12},
	Title = {Validation of the Urban Dispersion Model (UDM)},
	Volume = {8},
	Year = {2003}}

@article{legrand2009etl,
	Author = {Legrand, J. and Egan, J. R. and Hall, I. M. and Cauchemez, S. and Leach, S. and Ferguson, N. M.},
	Journal = {PLoS Computational Biology},
	Number = {1},
	Title = {Estimating the Location and Spatial Extent of a Covert Anthrax Release},
	Volume = {5},
	Year = {2009}}

@book{stc2011,
	Address = {Cambridge},
	Author = {{Shawe-Taylor}, J. and Cristianini, N.},
	Publisher = {Cambridge University Press},
	Title = {Kernel Methods for Pattern Analysis},
	Year = {2011}}

@book{thulasiraman1992gta,
	Address = {New York},
	Author = {Thulasiraman, K. and Swamy, M. N. S.},
	Publisher = {Wiley},
	Title = {Graphs: Theory and Algorithms},
	Year = {1992}}

@article{RESDSSE,
	Author = {Binois, M. and Huang, J. and Gramacy, R. B. and Ludkovski, M.},
	Journal = {Technometrics},
	Month = {October},
	Number = {1},
	Pages = {7-23},
	Title = {Replication or Exploration? Sequential Design for Stochastic Simulation Experiments},
	Volume = {61},
	Year = {2018}}

@article{jensen1998ait,
	Author = {Jensen, F. V.},
	Journal = {The Knowledge Engineering Review},
	Number = {2},
	Pages = {201-208},
	Title = {An Introduction to Bayesian Networks},
	Volume = {13},
	Year = {1998}}

@article{mchutchon2011gpt,
	Author = {McHutchon, A. and Rasmussen, C. E.},
	Journal = {Advances in Neural Information Processing Systems},
	Title = {Gaussian Process Training with Input Noise},
	Volume = {24},
	Year = {2011}}

@article{groer1978drc,
	Author = {Groer, P. G.},
	Journal = {Proceedings of the National Academy of Sciences of the United States of America},
	Number = {9},
	Pages = {4087-4091},
	Title = {Dose-Response Curves and Competing Risks},
	Volume = {75},
	Year = {1978}}

@article{titsias2010bgp,
	Author = {Titisias, M. K. and Lawrence, N. D.},
	Journal = {Proceedings of the 13th International Conference on Artificial Intelligence and Statistics},
	Pages = {844-851},
	Title = {Bayesian Gaussian Process Latent Variable Model},
	Volume = {9},
	Year = {2010}}

@article{gramacy2012btg,
	Author = {Gramacy, R. B. and Lee, H. K. H.},
	Journal = {Journal of the American Statistical Association},
	Number = {483},
	Pages = {1119-1130},
	Title = {Bayesian Treed Gaussian Process Models with an Application to Computer Modeling},
	Volume = {103},
	Year = {2012}}

@article{gramacy2015lgp,
	Author = {Gramacy, R. B. and Apley, D. W.},
	Journal = {Journal of Computational and Graphical Statistics},
	Number = {2},
	Pages = {561-578},
	Title = {Large Gaussian Process Approximation for Large Computer Experiments},
	Volume = {24},
	Year = {2015}}

@article{HDADGP,
	Author = {Dunlop, M. M. and Girolami, M. A. and Stuart, A. M. and Teckentrup, A. L.},
	Journal = {Journal of Machine Learning Research},
	Pages = {1-46},
	Title = {How Deep Are Deep Gaussian Processes},
	Volume = {19},
	Year = {2018}}

@article{DGP,
	Author = {Damianou, A. C. and Lawrence, N. D.},
	Journal = {Proceedings of the 16th International Conference on Artificial Intelligence and Statistics},
	Title = {Deep Gaussian Processes},
	Volume = {31},
	Year = {2013}}

@article{DVM,
	Author = {Chebyshev, P.},
	Journal = {Journal de math{\'e}matiques pures et appliqu{\'e}es},
	Number = {12},
	Pages = {177-184},
	Title = {Des valeurs moyennes},
	Volume = {2},
	Year = {1867}}

@book{LMRDMA,
	Address = {Boca Raton},
	Author = {Harville, D. A.},
	Publisher = {CRC press},
	Title = {Linear Models and the relevant distributions and matrix algebra},
	Year = {2018}}

@phdthesis{DPSEHMM,
	Author = {Jackson, S. E.},
	School = {Durham University},
	Title = {Design of Physical System Experiments Using Bayes Linear Emulation and History Matching Methodology with Application to Arabidopsis Thaliana},
	Year = {2018}}

@article{CMFP,
	Author = {Jha, B. and Juanes, R.},
	Journal = {Water Resources Research},
	Pages = {3776-3808},
	Title = {Coupled multiphase flow and poromechanics: A computational model of pore pressure effects on fault slip and earthquake triggering},
	Volume = {50},
	Year = {2014}}

@article{OC5,
	Author = {Taylor, K. E. and Stouffer, R. J. and Meehi, G. A.},
	Journal = {Journal of the American Meteorological Society},
	Pages = {485-498},
	Title = {An overview of CMIP5 and the Experiment design},
	Volume = {93},
	Year = {2012}}

@article{CSSCE,
	Author = {Loeppky, J. L. and Sacks, J. and Welch, W. J.},
	Journal = {Technometrics},
	Number = {4},
	Pages = {366-376},
	Title = {Choosing the Sample Size of a Computer Experiment: A Practical Guide},
	Volume = {51},
	Year = {2009}}

@article{PSEM,
	Author = {Allen, L. J. S.},
	Journal = {Infectious Disease Modelling},
	Pages = {128-142},
	Title = {A primer on stochastic epidemic models: Formulation, numerical simulation, and analysis},
	Volume = {2},
	Year = {2017}}

@article{CCMLSE,
	Author = {Kyzyurova, K. N. and Berger, J. O. and Wolpert, R. L.},
	Journal = {Journal on Uncertainty Quantification},
	Number = {3},
	Pages = {1151-1171},
	Title = {Coupling Computer Models Through Linking Their Statistical Emulators},
	Volume = {6},
	Year = {2018}}

@article{Jackson1,
	Author = {Jackson, S. E. and Vernon, I. and Liu, J. and Lindsey, K.},
	Journal = {Statistical Approaches in Genetics and Molecular Biology},
	Title = {Understanding Hormonal Crosstalk in Arabidopsis Root Development Via Emulation and History Matching},
	Volume = {19},
	Number = {5},
	Year = {2020}}

@misc{CE,
	Author = {Koehler, J. R. and Owen, A. B.},
	Title = {Computer Experiments},
	Year = {1996}}

@article{DPGP,
	Author = {Paulo, R.},
	Journal = {The Annals of Statistics},
	Number = {2},
	Pages = {556-582},
	Title = {Default priors for {G}aussian processes},
	Volume = {33},
	Year = {2005}}

@article{LBM,
	Author = {Hartigan, J. A.},
	Journal = {Journal of the Royal Statistical Society},
	Pages = {446-454},
	Title = {Linear {B}ayesian Methods},
	Volume = {31},
	Year = {1969}}

@article{BLERRM,
	Author = {O'Hagan, A.},
	Journal = {Journal of the American Statistical Association},
	Pages = {580-585},
	Title = {Bayes linear estimators for randomized response models},
	Volume = {82},
	Year = {1987}}

@book{PVE,
	Author = {Whittle, P.},
	Publisher = {Springer},
	Title = {Probability Via Expectation},
	Year = {1992}}

@book{Mardia79_MVA,
	Address = {London},
	Author = {Mardia, K. V. and Kent, J. T. and Bibby, J. M.},
	Publisher = {Academic Press},
	Title = {Multivariate Analysis},
	Year = 1979}

@incollection{AMA,
	Address = {Chichester},
	Author = {Goldstein, M. and Seheult, A. and Vernon, I.},
	Booktitle = {Environmental Modelling: Finding Simplicity in Complexity},
	Editor = {Wainwright, J. and Mulligan, M.},
	Publisher = {John Wiley and Sons},
	Title = {Assessing Model Adequacy},
	Year = {2013}}

@book{TP2,
	Author = {de Finetti, B.},
	Publisher = {Wiley},
	Title = {Theory of Probability},
	Volume = {2},
	Year = {1975}}

@article{3SR,
	Author = {Pukelsheim, F.},
	Journal = {The American Statistician},
	Number = {2},
	Pages = {88-91},
	Title = {The Three Sigma Rule},
	Volume = {48},
	Year = {1994}}

@article{DGPE,
	Author = {Bastos, T. S. and O'Hagan, A.},
	Journal = {Technometrics},
	Pages = {425-438},
	Title = {Diagnostics for {G}aussian process emulators.},
	Volume = {51},
	Year = {2008}}

@article{ENGPE,
	Author = {Andrianakis, Y. and Challenor, P. G.},
	Journal = {Computational Statistics and Data Analysis},
	Pages = {4215-4228},
	Title = {The Effect of the Nugget on {G}aussian Process Emulators of Computer Models},
	Volume = {56},
	Year = {2012}}

@article{PFTIMMPS,
	Author = {Goldstein, M. and Rougier, J. C.},
	Journal = {SIAM Journal on Scientific Computing},
	Number = {2},
	Pages = {467-487},
	Title = {Probabilistic Formulations for Transferring Inferences from Mathematical Models to Physical Systems},
	Volume = {26},
	Year = {2004}}

@book{TP,
	Author = {de Finetti, B.},
	Publisher = {Wiley},
	Title = {Theory of Probability},
	Volume = {1},
	Year = {1974}}

@incollection{BLA,
	Address = {New York},
	Author = {Goldstein, M.},
	Booktitle = {Encyclopedia of statistical Sciences},
	Chapter = {Bayes Linear Analysis},
	Editor = {Kotz, S. and Read, C. B. and Balakrishnan, N. and Vidakovic, B.},
	Pages = {29-34},
	Publisher = {Wiley},
	Title = {Bayes Linear Analysis},
	Year = {1999}}

@book{BLS,
	Address = {Chichester},
	Author = {Goldstein, M. and Wooff, D.},
	Publisher = {Wiley},
	Title = {Bayes Linear Statistics},
	Year = {2007}}

@article{BCCM,
	Author = {Kennedy, M. C. and O'Hagan, A.},
	Journal = {Journal of the Royal Statistical Society},
	Number = {3},
	Pages = {425-464},
	Title = {Bayesian Calibration of Computer Models},
	Volume = {63},
	Year = {2001}}

%%%%%%%%%%%%%%%%%%%%%%%%%%%%%%%%%%%%%%%%%%%%%%

\appendix

%%%%%%%%%%%%%%%%%%%%%%%%%%%%%%%%%%%%%%%%%%%%%%

\section{Bayes Linear Statistics \label{subsec:intro:BLS}}

In this article, we have largely focused on 
% the Bayes linear approach to analysis (and hence emulation - see Section \ref{subsec:intro:em}).
the Bayes Linear approach \citep{LBM, BLERRM, BLA, BLS} to statistical inference, which takes expectation as primitive, following De Finetti \citep{TP, TP2, PVE}, and deals with second-order belief specifications (that is, expectations, variances and covariances) of observable quantities.  Probabilities can be represented as the expectation of the corresponding indicator function when required.

More precisely, suppose that there are two collections of random quantities, $ \mb = (B_1,...,B_r) $ and $ \md = (D_1,...,D_s) $.  Bayes linear analysis involves updating subjective beliefs about $ \mb $ given observation of $ \md $.  In order to do so, prior mean vectors and covariance matrices for $ \mb $ and $ \md $ (that is, $ \expectation[\mb] $, $ \expectation[\md] $, $ \variance[\mb] $ and $ \variance[\md] $), along with a covariance matrix between $ \mb $ and $ \md $ (that is, $ \covariance[\mb,\md] $), must be specified.  Second-order beliefs about $ \mb $ can be adjusted in the light of $ \md $ using the Bayes linear update formulae:
\begin{eqnarray} 
\expectation_\md[\mb] & = & \expectation[\mb] + \covariance[\mb,\md]\variance[\md]^{-1}(\md-\expectation[\md]) \label{UE1} \\ \variance_\md[\mb] & = & \variance[\mb] - \covariance[\mb,\md]\variance[\md]^{-1}\covariance[\md,\mb] \label{UE2} \\
\covariance_\md[\mb_1, \mb_2] & = & \covariance[\mb_1, \mb_2] - \covariance[\mb_1,\md]\variance[\md]^{-1}\covariance[\md,\mb_2] \label{UE3}
\end{eqnarray}  
\sam{Equations \eqref{UE1}-\eqref{UE3} are the backbone of the Bayes linear update Equations (4) and (5) of Section 2.1 of the main text.}
$ \expectation_\md[\mb] $ and $ \variance_\md[\mb] $ are termed the adjusted expectation and variance of $ \mb $ given $ \md $ \citep{BLS}.  $ \covariance_\md[\mb_1, \mb_2] $ is termed the adjusted covariance of $ \mb_1 $ and $ \mb_2 $ given $ \md $, where $ \mb_1 $ and $ \mb_2 $ are subcollections of $ \mb $.   
Following on from this, given a third collection of random quantities $ \ma = (A_1,...,A_t) $ we can sequentially adjust beliefs about $ \mb $ given observation of random quantities $ \md $ and $ \ma $ using a sequential Bayes linear adjustment:
\begin{eqnarray} 
\expectation_{\md \cup \ma}[\mb] & = & \expectation_{\md}[\mb] + \covariance_\md[\mb,\ma]\variance_\md[\ma]^{-1}(\ma-\expectation_\md[\ma]) \label{UES1} \\ \variance_{\md \cup \ma}[\mb] & = & \variance_\md[\mb] - \covariance_\md[\mb,\ma]\variance_\md[\ma]^{-1}\covariance_\md[\ma,\mb] \label{UES2} 
\end{eqnarray}
which adjusts the adjusted beliefs of $ \mb $ by $ \md $ now additionally by $ \ma $.
Note that equivalent results are obtained by updating first by $ \ma $ then $ \md $ by swapping the occurrences of $ \md $ and $ \ma $ in Equations \eqref{UE1}-\eqref{UES2} above.
Equations \eqref{UES1} and \eqref{UES2} are important for some of the discussions and calculations presented throughout Section 3 of the main text.
%For a more detailed overview of Bayes linear methods, see \cite{BLA}, and for a thorough treatment, see \cite{BLS_App}.
%For a comparison of Bayes linear methods with the full Bayesian approach, see, for example, \cite{BLS_App, BLA} and \cite{GFBUA}.

%%%%%%%%%%%%%%%%%%%%%%%%%%%%%%%%%%%%%%%%%%%%%%

\section{Proof of \sam{Lemmas \ref{pdlemma}-\ref{lem2}} \label{LemmaProofs}}

In this section, we prove Lemmas 1 - 4
of the main text.

%%%%%%%%%%%%%%%%%%%%%%%%%%%%%%%%%

\subsection*{Proof of Lemma 3.2.1}

Rewrite Equation (21) as
\be \nonumber
\begin{split}
c(\bX,\bX') & = \exp \left\{ - \e{(\bX - \bX')^T\Theta^{-2}(\bX - \bX')}\right\}\\
&= \exp \left\{ - \e{\bX^T\Theta^{-2}\bX} \right\} \exp \left\{ -\e{(\bX')^T\Theta^{-2}\bX'} \right\} \exp \left\{ 2\e{\bX^T\Theta^{-2}\bX'} \right\}\,, \\
\end{split}
\ee
with kernel $$ k(\bX , \bX') = \exp \left\{ 2\e{\bX^T\Theta^{-2}\bX'} \right\} \, . $$
The kernel $ k(\bX, \bX') $ is positive definite since 
$$ \mathrm{tr}(\Theta^{-2}\var{\bX}) + \e{\bX}^T\Theta^{-2}\e{\bX} \ge 0 $$ 
with equality if and only if $\bX$ has a degenerate distribution at zero. Postive definiteness of the function $c(\bX, \bX')$ then follows from standard properties of kernels (see \citealp{stc2011}, ch.~3)

%%%%%%%%%%%%%%%%%%%%%%%%%%%%%%%%%

\subsection*{Proof of Lemma 3.2.2}

Covariance can be derived from conditional quantities using the law of total covariance, hence:
\begin{multline*}
\cov{\uX}{\uXp} = \e{\cov{\uX}{\uXp\,|\,\bX = \bx, \bX'=\bx'}} \\ 
+ \cov{\e{\uX\,|\, \bX}}{\e{\uXp\,|\, \bX'}}\,.
\end{multline*}
Under the assumption that $\e{\bu(\bX)} = 0$, it follows that $\cov{\e{\uX\,|\, \bX}}{\e{\uXp\,|\, \bX'}} = 0$. Hence for conditional covariance (22), we have
\begin{equation*}
\begin{split}
\cov{\uX}{\uXp} & = \mathrm{E}\left[
\exp \left( - \sum_{r=1}^p \left\{ \frac{X_{(r)} - X'_{(r)}}{\theta_r} \right\}^2  \right)
\right] \, \Sig \\
& \succeq  \exp \left( 
- \sum_{r=1}^p \mathrm{E}\left[
\left\{ \frac{X_{(r)} - X'_{(r)}}{\theta_r} \right\}^2 \,
\right]
\right) \, \Sig \, , \\
\end{split}
\end{equation*}
under the Loewner (partial) ordering, following from an application of Jensen's inequality and the positive-definiteness of $\Sig$. As the diagonal entries of $\Sig$ are all non-negative, elementwise inequality for the $kk$th entry follows directly ($k = 1, \ldots, q$).

%%%%%%%%%%%%%%%%%%%%%%%%%%%%%%%%%

\subsection*{Proof of Lemma 3.2.3}

\sam{We begin by expanding the terms of the Bayes linear update as follows:}
\ba
\eF{\fX} & = &  \eF{\wX \, \bb}   +   \eF{\uX}\,.  \nonumber
\ea
Taking the two parts of the right-hand side of this equation separately, we first have that
\be
\eF{\wX \, \bb} = \eF{\wX} \, \eF{\bb} = \e{\wX} \, \eF{\bb}\,,  \label{eFwXb} 
\ee
where we have used the facts that $ \eF{\wX} = \e{\wX} $ since $ \cov{\bX}{\bF} = \bzero $, and $ \covF{\wX}{\bb} = \bzero $ since $ \covF{\gX}{\bb} = \bzero $.
\sam{We then have, using basic rules of linear algebra \citep{Mardia79_MVA}, that
\ba
\eF{\uX} & = & \e{\uX} + \cov{\uX}{\bF} \, \var{\bF}^{-1} \, ( \bF - \e{\bF} )
\NLeq
\cov{\uX}{\bU} \, \var{\bF}^{-1} \, (\bF - \bW \, \bGa )
\NLeq
\vX \, ( \WDW + \bO )^{-1} \, ( \bF - \bW \, \bGa )
\NLeq
\vX \, \big( ( \WDW + \bO )^{-1} \, \bF - ( \WDW + \bO )^{-1} \, \bW \, \bGa \big)
\NLeq
\vX \bigg( \big( \bOi - \bOi \, \bW \, ( \bDi + \WOW )^{-1} \, \bW^T \, \bOi \big) \, \bF 
\NL
- \bOi \, \bW \, ( \WOW + \bDi )^{-1} \, \bDi \, \bW \bigg)  
\NLeq
\vX \, \bOi \, \big( \bF - \bW \, ( \bDi + \WOW )^{-1} \, ( \bDi \, \bW + \bW \, \bOi \, \bF ) \big) 
\NLeq
\vX \, \bOi \, ( \bF - \bW \, \eF{\bb} )\,.   \label{eFuX}
\ea
Combining Equations \eqref{eFwXb} and \eqref{eFuX} we get:
}
\ba
\eF{\fX} & = & \ewX \, \eb   +    \vX \, \bOi \, (\bF - \bW \, \eb )\,. \nonumber
\ea
\begin{flushright} $ \Box $ \end{flushright}

%%%%%%%%%%%%%%%%%%%%%%%%%%%%%%%%%

\subsection*{Proof of Lemma 3.2.4}

\sam{We begin by expanding the terms of the Bayes linear update as follows:}
\ba
\varF{\fX} & = & \varF{\wX \, \bb + \uX}   
\NLeq
\varF{\wX \, \bb} + \varF{\uX}  
\NL 
\, + \, \covF{\wX \, \bb}{\uX} + \covF{\uX}{\wX \, \bb}\,.    \label{var_gZ} 
\ea
%%%
\sam{Taking each term on the right-hand side of Equation \eqref{var_gZ} in turn, we have, using linear algebra \citep{Mardia79_MVA}:}
\ba
\varF{ \wX \, \bb } & = & \e{ \varFX{ \wX \, \bb } } + \var{ \eFX{ \wX \, \bb } }
\NLeq
\e{ \wX \, \vb \, \wX^T }    +    \ebt \, \var{ \wX }  \eb \,,   \label{varFwXb}
\ea
%%%
\sam{
\ba
\lefteqn{ \varF{ \uX } } \nonumber \\ 
& = &  \var{ \uX } -  \cov{\uX}{\bF} \, \var{\bF}^{-1} \, \cov{\bF}{\uX}
\NLeq
\Sig - \vX \, ( \WDW + \bO )^{-1} \, \vX^T
\NLeq
\Sig - \vX \, ( \bOi - \bOi \, \bW \, ( \bDi + \WOW )^{-1} \, \bW^T \, \bOi ) \, \vX^T
\NLeq
\Sig - \vX \, \bOi \, \vX^T   +   \vX \, \bOi \, \bW \, \vb  \, \bW^T \, \bOi \, \vX^T \,,  \label{varFuX}
\ea
}
and
%%%
\sam{
\ba
\lefteqn{\covF{ \wX \, \bb }{ \uX }} \nonumber \\ 
& = & \cov{ \wX \, \bb }{\uX} - \cov{ \wX \, \bb }{\bF} \var{\bF}^{-1} \cov{\bF}{\uX} \, . \label{thirteen}
\ea
In Equation \eqref{thirteen}, we have that:
\ba
\cov{ \wX \, \bb }{\bF} & = & \cov{ \wX \, \bb}{ \bW \,\bb}
\NLeq
\cov{ \wX \, \bb}{ \bb} \bW^T
\NLeq
( \e{ \wX \, \bb \, \bb^T } - \e{ \wX \, \bb } \e{ \bb^T } ) \bW^T
\NLeq
( \e{ \wX } \e{ \bb \, \bb^T } - \e{ \wX } \e{ \bb } \e{ \bb^T } ) \bW^T
\NLeq
\e{ \wX } \var{\bb} \bW^T
\NLeq
\e{ \wX } \bDe \bW^T\,, \nonumber
\ea
}
so that
\ba
\covF{ \wX \, \bb}{\uX} & = & - \, \e{\wX} \, \bDe \, \bW^T (\WDW + \bO)^{-1} \vX^T
\NLeq
- \, \e{\wX} ( \WOW + \bDi )^{-1} \bW \, \bOi \, \vX^T
\NLeq
- \, \e{\wX} \vb \WO \vX^T \,.  \label{fiftenn}
\ea
\sam{Putting Equations \eqref{varFwXb}, \eqref{varFuX} and \eqref{fiftenn} together, we get that:}
\ba
\varF{ \fX } & = & \e{ \wX \, \vb \, \wX^T }   +   \ebt \, \vwX \, \eb + \, \Sig  
\NL
\,    - \,   \vX \, \bOi \, \vX^T    +    \vX \, \bOi \, \bW \, \vb  \, \bW^T \, \bOi \, \vX^T
\NL
\, - \, \ewX \, \vb \, \bW \, \bOi \, \vX^T    
\NL
\, - \,   ( \ewX \, \vb \, \bW \, \bOi \, \vX^T ) ^T  \label{UIBLE_lem_Var}
\ea
\begin{flushright} $ \Box $ \end{flushright}

%%%%%%%%%%%%%%%%%%%%%%%%%%%%%%%%%%%%%%%%%%%%%%

\vspace{-0.3cm}

\section{Extension of the Networks Example \label{ExtNetEg}}

\vspace{-0.1cm}

\begin{figure}
\centering
\includegraphics[height=17cm, width = 14cm, trim = {0, 3cm, 0, 3cm}]{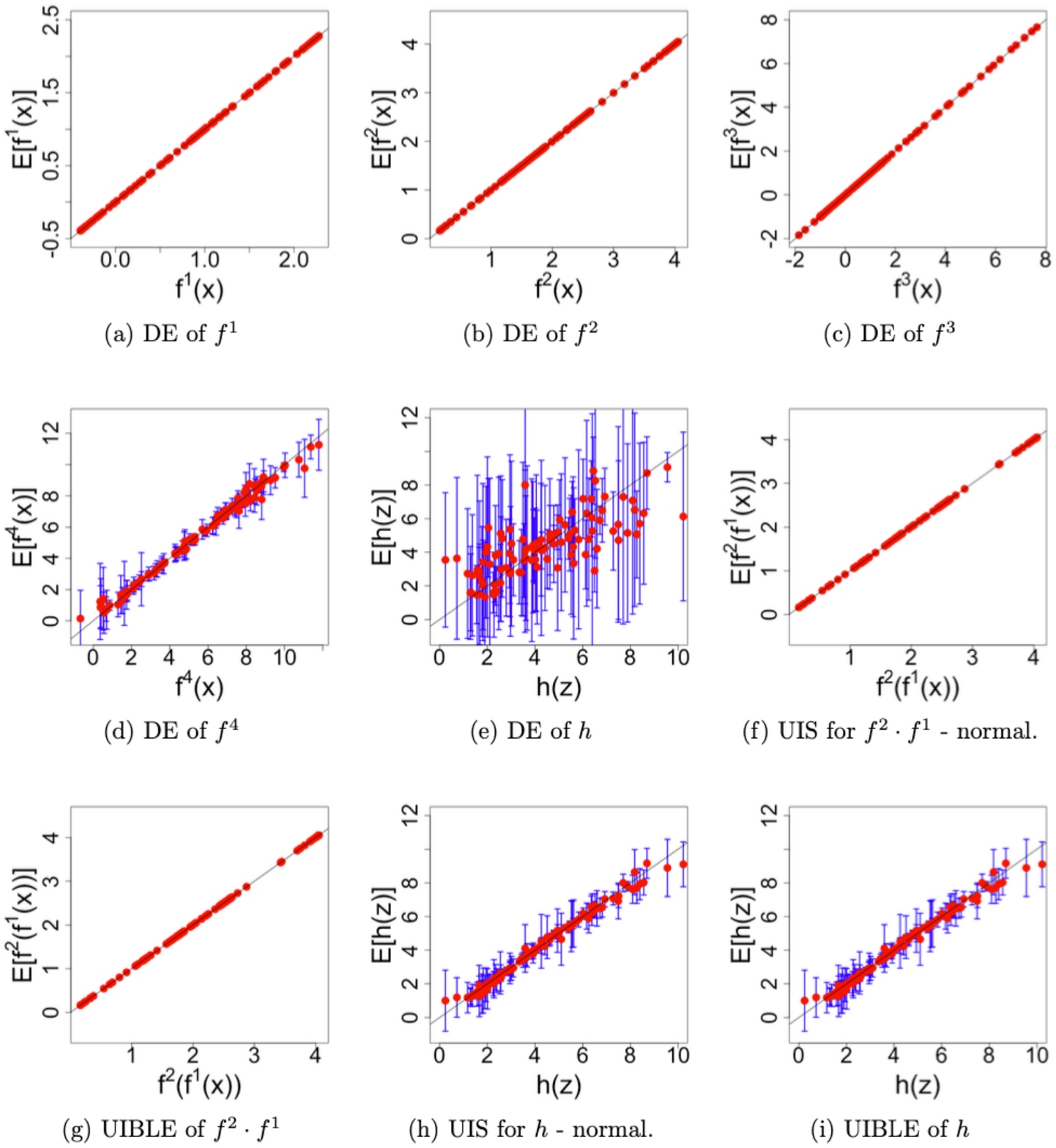}
\caption{Adjusted expectation $ \pm 3 $ standard deviations against simulator output for \sam{nine emulators discussed in the text.}}
\label{NetworkExampleApp1}
\end{figure}

Figure \ref{NetworkExampleApp1} shows diagnostic plots for \sam{nine emulators relating to the larger simulator network example Section 5 of the main text, using 30 training points to train each emulator.}
\sam{Figures \ref{NetworkExampleApp1}a-\ref{NetworkExampleApp1}e show diagnostic plots for DE of $ f^1, f^2, f^3, f^4 $ and $ h $.}
Since $ f^1 $, $ f^2 $ and $ f^3 $ are relatively simple 1-dimensional functions, 30 training points allow almost-perfect predictions.  
$ f^4 $ is emulated \sam{with some uncertainty but fairly accurately}, however, it is difficult to mimic the behaviour of $ h $ using DE.  
\sam{Figures \ref{NetworkExampleApp1}f and \ref{NetworkExampleApp1}g show diagnostic plots for the approximation of $ f^2(f^1(\cdot)) $ using UIS and UIBLE.
Both result in highly accurate and precise emulators as a result of the accuracy and precision of the component emulators for $ f^1 $ and $ f^2 $ (diagnostics shown in Figures \ref{NetworkExampleApp1}a and \ref{NetworkExampleApp1}b).
Figures \ref{NetworkExampleApp1}h and \ref{NetworkExampleApp1}i show diagnostic plots for the approximation of $ h $ using UIS and UIBLE, these being identical to Figures 5b and 5c of the main text, but shown again here for comparison purposes.}

\begin{figure}
\centering
\includegraphics[height=17cm, width = 14cm, trim = {0, 3cm, 0, 3cm}]{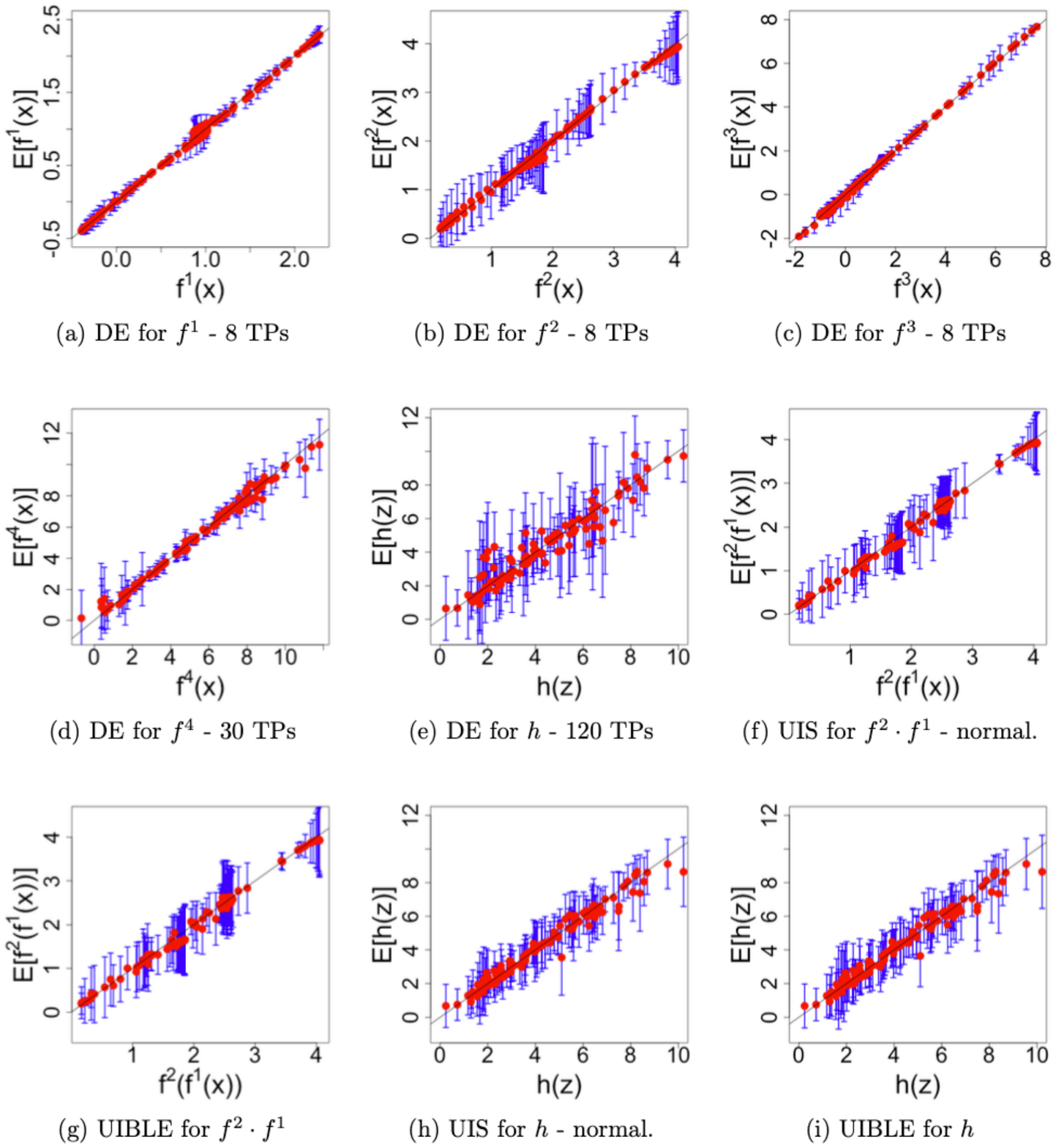}
\caption{Adjusted expectation $ \pm 3 $ standard deviations against simulator output for \sam{nine} emulators \sam{discussed in the text.}}
\label{NetworkExampleApp2}
\end{figure}

Figure \ref{NetworkExampleApp2} shows \sam{nine corresponding} diagnostic plots for the case of reducing the number of 
training points for the emulators of the one-dimensional simulators $ f^1 $, $ f^2 $ and $ f^3 $ to 8, 
whilst increasing the number for $ h $ to 120.  
Whilst $ f^1 $ and $ f^3 $ still have fairly low uncertainty, the uncertainty on $ f^2 $ is higher, though all three emulators have high accuracy.  
As discussed in the main text, DE for $ h $ constructed using 120 training points is much more accurate, although only similarly accurate to UIS and UIBLE (Figures \ref{NetworkExampleApp2}h and \ref{NetworkExampleApp2}i) using many fewer points.
\sam{The uncertainties in the approximation of $ f^2(f^1(\cdot)) $ using UIS and UIBLE reflect accurate and fairly precise predictions, and are in accordance with the alternative 1-dimensional diagnostic plots shown in Figures 3d and 3f of the main text.}


\end{document}